\documentclass[12pt]{article}

\usepackage{scicite}
\usepackage{amsmath}
\usepackage{times}
\usepackage{graphicx}
\usepackage{braket}
\usepackage{color}
\usepackage{siunitx}

\newenvironment{sciabstract}{%
\begin{quote} \bf}
{\end{quote}}

\title{
Nonlinear feedforward enabling quantum computation
}

\author
{Atsushi Sakaguchi,$^{1,2,\ast}$
 Shunya Konno,$^{1}$
 Fumiya Hanamura,$^{1}$\\
 Warit Asavanant,$^{1,2}$
 Kan Takase,$^{1,2}$
 Hisashi Ogawa,$^{1}$
 Petr Marek,$^{3}$
 Radim Filip,$^{3}$\\
 Jun-ichi Yoshikawa,$^{1,2}$
 Elanor Huntington,$^{4}$
 Hidehiro Yonezawa,$^{5}$
 Akira Furusawa$^{1,2,\ast\ast}$\\
\\
\normalsize{$^{1}$Department of Applied Physics, School of Engineering, The University of Tokyo,}\\
\normalsize{7-3-1 Hongo, Bunkyo-ku, Tokyo 113-8656, Japan,}\\
\normalsize{$^{2}$Optical Quantum Computing Research Team, RIKEN Center for Quantum Computing,}\\
\normalsize{2-1 Hirosawa, Wako, Saitama, 351-0198, Japan}\\
\normalsize{$^{3}$Department of Optics, Palack\'y University,}\\
\normalsize{17. listopadu 1192/12, 77146 Olomouc, Czech Republic}\\
\normalsize{$^{4}$Centre for Quantum Computation and Communication Technology and }\\
\normalsize{Energy
and Materials Engineering, College of Engineering and Computer Science,}\\
\normalsize{Australian National University, Canberra, ACT 2600, Australia}\\
\normalsize{$^{5}$Centre for Quantum Computation and Communication Technology,}\\
\normalsize{School of Engineering and Information Technology, }\\
\normalsize{University of New South Wales, Canberra, ACT 2600, Australia.}\\
\normalsize{$^\ast$To whom correspondence should be addressed; E-mail:  atsushi.sakaguchi@riken.jp.}\\
\normalsize{$^{\ast\ast}$ E-mail:  akiraf@ap.t.u-tokyo.ac.jp.}
}

\date{}

\begin{document}


\baselineskip24pt


\maketitle


\begin{sciabstract}
Measurement-based quantum computation with optical time-domain multiplexing is a promising method to realize a quantum computer from the viewpoint of scalability.
Fault tolerance and universality are also realizable by preparing appropriate resource quantum states and electro-optical feedforward that is altered based on measurement results. %from the measurement to quantum states.
% In this method, the variations of the measurements determine the universality.
% To realize universal measurement-based quantum computer, both linear and nonlinear quadrature measurement are required.
While a linear feedforward has been realized and become a common experimental technique, nonlinear feedforward was unrealized until now. %until now, unrealized due to technological difficulties.
In this paper, we demonstrate that a fast and flexible nonlinear feedforward realizes the essential measurement required for fault-tolerant and universal quantum computation. Using non-Gaussian ancillary states we observed  10$\%$  reduction of the measurement excess noise relative to classical vacuum ancilla. % , with excess noise reduced by 10$\%$ relative to the vacuum level.
%Furthermore, by combining this nonlinear feedforward with non-Gaussian ancillary states, we realize nonlinear quadrature measurement in an optical system required for universality.
% The measurement is compatible with measurement-based quantum computation.
% We develop a flexible and fast nonlinear feedforward technique and combined it with the non-Gaussian ancillary states and the homodyne measurements, making our measurement compatible with the measurement-based quantum computation.
%The quantum non-Gaussianity of the tailored measurement is evaluated via detector tomography which shows the success of our measurement.
%Nonlinear feedforward demonstrated in this work has a wide range of applications including quantum error correction, and will be an imperative technology for realizing a quantum computer.
% The established nonlinear feedforward technique enables various operations such as error correction and forms an essential technology in the realization of the optical quantum computer.
\end{sciabstract}

\begin{figure}
    \centering
    \includegraphics[width=.9\textwidth]{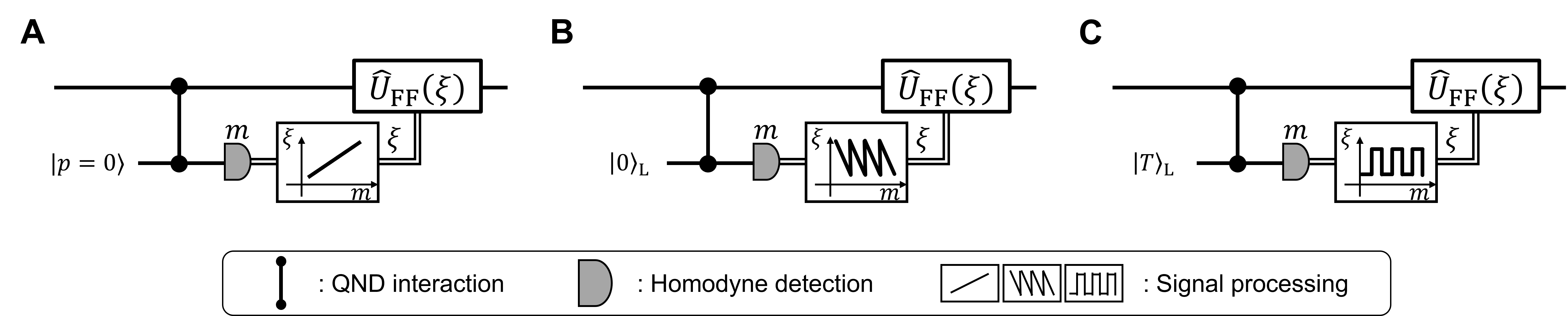}
    \caption{\textbf{Quantum operations implemented in measurement-based quantum computation.} (A) Clifford (Gaussian) operation via linear feedforward and a Gaussian ancillary state. (B) Error recovery operation of GKP encoding via nonlinear feedforward and a GKP logical state as an ancilla. (C) Gate teleportation scheme for a fault-tolerant non-Clifford operation via nonlinear feedforward and a magic ancillary state. QND : quantum non-demolition.
    }
    \label{fig1}
\end{figure}

% In setting up this template for *Science* papers, we've used both
% the \section* command and the \paragraph* command for topical
% divisions.  Which you use will of course depend on the type of paper
% you're writing.  Review Articles tend to have displayed headings, for
% which \section* is more appropriate; Research Articles, when they have
% formal topical divisions at all, tend to signal them with bold text
% that runs into the paragraph, for which \paragraph* is the right
% choice.  Either way, use the asterisk (*) modifier, as shown, to
%Quantum computation is one of the most focused research topics in last decades,
% \color{red}
% Quantum computers are expected to enable drastically faster computation than classical computers in specific tasks and has been one of the most focused research topics in last decades.
A quantum computer promises to solve certain computational tasks significantly faster than a modern computer does. Nowadays, quantum computing is one of the hottest research topics.
The goal of the research is to realize a practical quantum computer that is scalable, universal and fault tolerant.
Many physical systems\cite{doi:10.1126/science.abb2823}---e.g., superconducting devices\cite{Nakamura1999, Jurcevic_2021}, trapped-ion systems\cite{PhysRevLett.60.535, PhysRevLett.113.220501}, and semiconductor systems\cite{PhysRevA.57.120,Veldhorst2014}---have been investigated extensively.
Among them, optical systems have unique potential regarding the scalability \cite{kashiwazaki2020,kashiwazaki2021,https://doi.org/10.48550/arxiv.2205.14061,endo2021}.
For example, generation of a large-scale entangled states, the so-called cluster states, has been demonstrated using time-domain multiplexing methods\cite{doi:10.1063/1.4962732, doi:10.1126/science.aay2645, doi:10.1126/science.aay4354}.
The cluster states are resources of measurement-based quantum computation (MBQC), where quantum operations are performed via local measurements on the large-scale cluster states and feedforward operations depending on the measurement outcomes\cite{PhysRevLett.86.5188, PhysRevA.73.032318}.
The demonstrated large-scale cluster states are categorized as continuous-variable cluster states, which treats continuous-valued quadratures $\hat{x}$ and $\hat{p}$ of an electro-magnetic field that satisfy $\left[\hat{x}, \hat{p}\right] = i\hbar$.
In continuous-variable MBQC, homodyne measurement is one of the most fundamental and powerful  measurement\cite{RevModPhys.77.513}. When combined with ancillary states and feedforward, homodyne measurement has an ability to implement fault-tolerant universal quantum computation\cite{PhysRevResearch.3.043026, PhysRevA.64.012310, PhysRevLett.119.180507}.
For example, this combination can implement Clifford operations or Gaussian operations (Fig.\ref{fig1}A), error recovery operations (Fig.\ref{fig1}B), and fault-tolerant non-Clifford operations (Fig.\ref{fig1}C).
It is, however, emphasized that ancillary states and feedforward must be specifically customized for each operations\cite{PhysRevResearch.3.043026}.
In previous research, only deterministic Gaussian operations on Gaussian and non-Gaussian states have been demonstrated with Gaussian ancillary states and solely linear feedforward\cite{PhysRevApplied.16.034005, Larsen2021, PhysRevLett.113.013601}.
Deterministic non-Gaussian operations have not been realized so far.

The difficulty of implementing deterministic non-Gaussian operations on traveling optical states stems from the requirement of complicated non-Gaussian ancillae and nonlinear feedforward --- conditional Gaussian operations controlled by the nonlinear function of the measurement outcomes.
While the preparation of ancillary states in optical systems has been extensively researched theoretically\cite{Glancy:08, Vasconcelos:10, PhysRevA.97.022341, PhysRevLett.123.200502} and experimentally\cite{Hacker2019, Arrazola2021}, the development of essential nonlinear feedforward has remained limited.
There are a few reports about nonlinear feedforward such as digital feedforward with a primitive digital logic\cite{Prevedel2007} or an analog feedforward with dedicated circuits for a specific task\cite{PhysRevA.90.060302}.
The nonlinear feedforward in the previous researches is, however, inflexible or too slow to synchronize electrical signals and optical signals, which is imperative for MBQC in time domain.
Hence, nonlinear feedforward is a key piece to unlock the full potential of an optical quantum computer.

\begin{figure}
    \centering
    \includegraphics[width=0.9\textwidth]{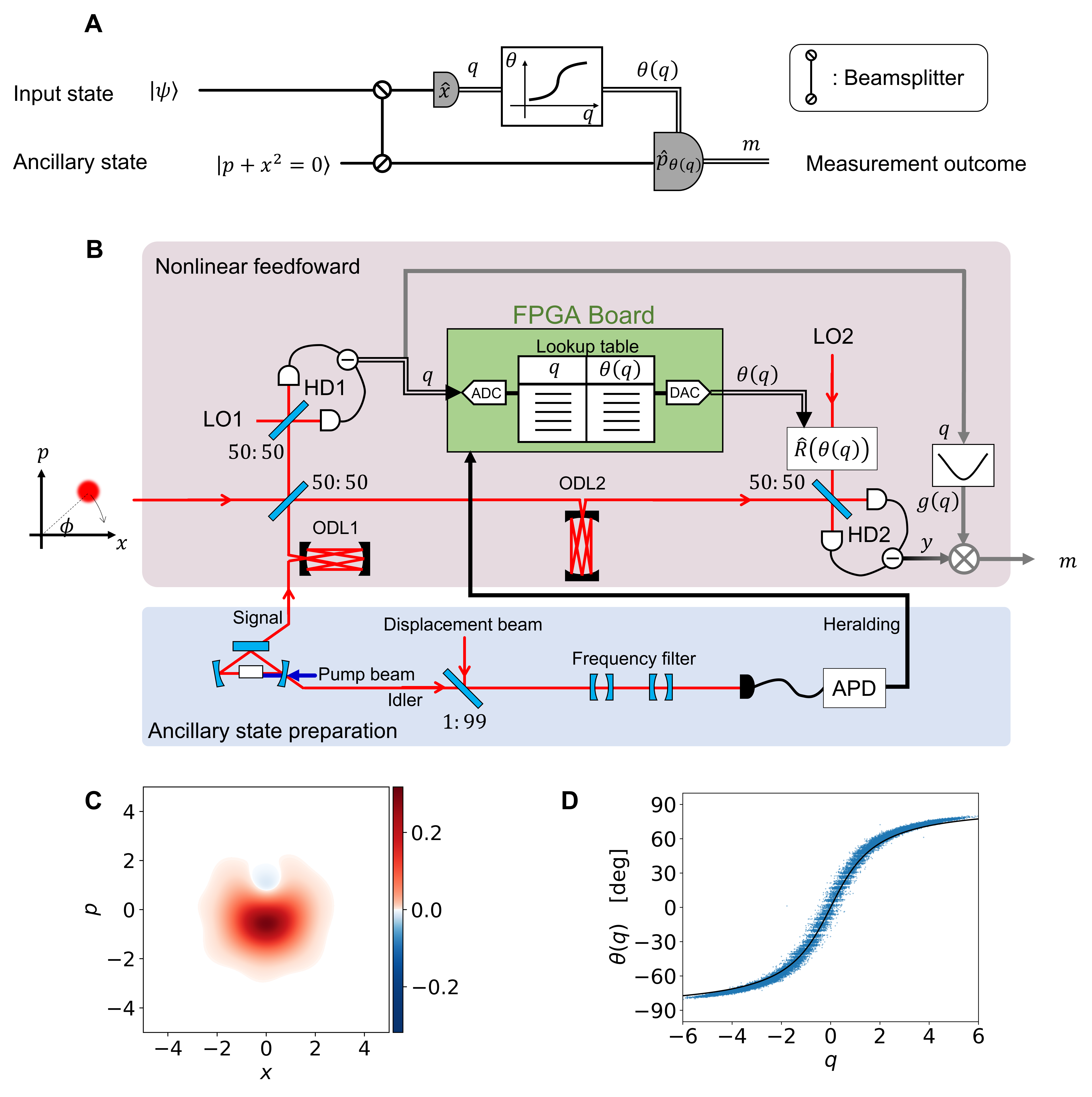}
    \caption{\textbf{Nonlinear quadrature measurement setup}.
    (A) Quantum circuit diagram of the nonlinear quadrature measurement.
    (B) Schematic of the experimental setup. Input states and ancillary states are localized in time domain.
    The whole system consists of two parts, a setup for ancillary state preparation via heralding method and a setup for nonlinear feedforward operation.
    In the ancillary state preparation, an optical parametric oscillator (OPO) is pumped by a frequency doubled beam, generating a two-mode squeezed state.
    One of the mode (idler mode) is displaced and passed through frequency filters before it is measured by an avalanche photodiode (APD).
    Click events by the APD herald the success of non-Gaussian ancillary states preparation in the signal mode and trigger the nonlinear feedforward operations.
    See the supplemental material for details of the experimental setup. LO : Local oscillator, HD : homodyne detector, ODL : optical delay line, FPGA : Field programmable gate array, ADC : Analog-to-digital converter, DAC : Digital-to-analog converter, APD : Avalanche photodiode.
    (C) Wigner function of the ancillary state used in this experiment. % and its cross section along $x=0$. The negative region and parabolic shape of Wigner function are the signs of the quantum non-Gaussianity of the ancillary state.
    % \textbf{Note: Remove cross section of the Wigner function.}
    (D) Input-output relation of the nonlinear calculation on the FPGA board. The target function, $\theta(q) = \arctan (\sqrt{2}\gamma q)$ where $\gamma=0.52$, is shown as a black line. Blue dots show experimentally obtained values.}
    \label{fig2}
\end{figure}

Here, we demonstrate flexible and fast nonlinear electro-optical feedforward and use it to implement a nonlinear quadrature measurement that, in combination with a suitable ancilla, projects the state of traveling light into a non-Gaussian state, as is required for quantum computing.
%with a non-Gaussian ancillary state.
Our setup (Fig.\ref{fig2}A) measures the nonlinear combination of two quadratures $\hat{x}$ and $\hat{p}$ of an electromagnetic field, $\hat{p}+\gamma\hat{x}^2$ where $\gamma$ is a parameter we can tune.
This nonlinear quadrature measurement can be readily applied to a non-Clifford operation if combined with the cluster states already demonstrated in \cite{doi:10.1126/science.aay2645, doi:10.1126/science.aay4354, PhysRevApplied.16.034005, Larsen2021}.
We perform tomography of the tailored measurement, observe 10 $\%$ reduction of excess noise thanks to the non-Gaussian ancilla, and verify its quantum non-Gaussian nature.
The results signify that our feedforward system works properly and the nonlinear quadratures are indeed measured.
The nonlinear feedforward system developed in this work is flexible and capable of implementing various signal processing.
Therefore, it is applicable not only to specific non-Gaussian operations, but also to fault-tolerant non-Clifford operations on GKP qubits\cite{PhysRevResearch.3.043026}, continuous-variable gate teleportation\cite{PhysRevA.64.012310} and analog error correction of GKP qubits\cite{PhysRevA.64.012310, PhysRevLett.119.180507}, if appropriate ancillary states are prepared.
This work has opened a new nonlinear regime beyond large-scale cluster states and Gaussian operations, establishing an important cornerstone of optical quantum computation.
\color{black}

Figure \ref{fig2}B shows a schematic diagram of the experimental system.
An input state is interfered with an ancillary state on a beam splitter, and one of the outputs is measured by a homodyne detector (HD1).
The measured value $q$ of the homodyne detector is processed by nonlinear functions on a field programmable gate array (FPGA) board.
During this signal processing, the other output of the beam splitter is on hold in an optical delay line.
The calculated results by the FPGA are fed forward to the other homodyne detector (HD2) and set the measurement basis  to $\hat{p}_{\theta(q)} = \hat{p}\,\cos\left(\theta\left(q\right)\right) + \hat{x}\,\sin \left(\theta\left(q\right)\right)$.
The exact form of the nonlinear feedforward is determined by $\theta ( q )=\arctan \left( \sqrt{2} \gamma q \right)$ as shown in Fig. \ref{fig2}D.
To measure the nonlinear quadrature of $\hat{p}_{\mathrm{in}}+\gamma\hat{x}^2_{\mathrm{in}}$, the measurement outcome of the second homodyne detector, $y$, is multiplied by the gain
$\sqrt{2}\slash\cos\left(\theta \left(q\right)\right)$ which is determined by the measurement outcome $q$ of the first homodyne detector.
Finally, the outcome $m=\sqrt{2}\,y\slash \cos\left(\theta\left(q\right)\right)$ is obtained.
This $m$ corresponds to the nonlinear quadrature $\hat{m}$,
\begin{align}
\hat{m} =
\hat{p}_{\mathrm{in}} +\gamma\hat{x}^{2}_{\mathrm{in}} +
\left( \hat{p}_{\mathrm{anc}} -\gamma\hat{x}^{2}_{\mathrm{anc}} \right)
\label{eq1}
\end{align}
where $\hat{x}_{\mathrm{in}}, \hat{p}_{\mathrm{in}}$ are quadratures of the input state, and $\hat{x}_{\mathrm{anc}}, \hat{p}_{\mathrm{anc}}$ are quadratures of the ancillary state.
Equation \eqref{eq1} shows that the nonlinear quadrature of the input state, $\hat{\delta}_{\mathrm{in}} = \hat{p}_{\mathrm{in}}+\gamma\hat{x}^2_{\mathrm{in}}$, is influenced by an excess noise caused by the corresponding nonlinear quadrature of the ancillary state, $\hat{\delta}_{\mathrm{anc}} = \hat{p}_{\mathrm{anc}} -\gamma\hat{x}^{2}_{\mathrm{anc}}$.
Note that the excess noise is independent from quadratures of the input state. Hence, the amount of excess noise is determined only by the ancillary state.
The ideal ancillary state that gives $\delta_\mathrm{anc = 0}$ is a cubic phase state (CPS), which satisfies
\begin{align}
    \hat{\delta}_{\mathrm{anc}} \ket{\mathrm{CPS}} = 0.
    \label{eq2}
\end{align}
An ideal CPS is an unphysical state because it requires infinite energy to generate. Thus, we must consider an approximated CPS similar to squeezed states substituted for ideal quadrature eigenstates in continuous-variable quantum computation.
We call the approximated cubic phase state as a nonlinearly squeezed state or cubic squeezed state\cite{Kala:22}, since the variance of the nonlinear quadrature operator $\hat{\delta}_{\mathrm{anc}}$ is squeezed beyond the lower bound imposed by Gaussian states and their mixtures\cite{PhysRevApplied.15.024024}.
It is known that a superposition of photon number states can be a good approximation of nonlinearly squeezed state even with a moderate number of photons in the state \cite{PhysRevA.88.053816, PhysRevApplied.15.024024}.
Figure \ref{fig2}C shows the Wigner function of the ancillary state used in our experiment. This ancillary state is nonlinearly squeezed by about 10\% beyond any Gaussian states or their mixtures when $\gamma=0.52$ and it has clear regions of negativity.
Note that the level of the nonlinear squeezing depends on the coefficient $\gamma$, and this value is optimal for the experimental ancillary state.
The ancillary state can be further improved by increasing the number of the photons to generate larger nonlinear squeezing\cite{PhysRevA.64.012310, PhysRevLett.124.240503}.

Fast nonlinear feedforward system is a key technological component in our challenging experimental setup.
This is because slow signal processing leads to a long optical delay line which entails adverse effects such as loss and phase fluctuation. To implement fast and flexible signal processing, we used an FPGA board equipped with low-latency AD/DA converters \cite{Supplement} and implement a look-up table inside the FPGA.
The target function (arctangent) is pre-calculated and stored in the look-up table so that
the calculation is accurately completed within 1 clock cycle, 2.67\,ns in this experiment. The total latency of the FPGA board is 26.8\,ns, corresponds to about 8 meters of optical delay lines which can be feasibly stabilized in experimental setups.
This latency does not depend on the form of processing as the values of look-up tables are calculated in advance, thus the feedforward system has significant flexibility.

To experimentally characterize our nonlinear measurement, we input various coherent states $\ket{\alpha}$, where $\alpha$ is the complex amplitude written by two real values $\alpha_x, \alpha_p$ as $\alpha = \left(\alpha_x +i\alpha_p \right) \slash \sqrt{2}$.
The input coherent states are carefully calibrated by dual homodyne measurements\cite{RevModPhys.77.513}, which is implemented by the same experimental setup with the feedforward system turned off.
We choose 27 different amplitudes $|\alpha|$ equally spaced and ranging from 0 to 3.5.
The coherent states are sampled with randomized phase in each fixed amplitude $|\alpha|$\cite{Supplement}.

In the Heisenberg picture, the quality of the measurement can be analyzed by looking at the first and second moments of the measured nonlinear quadrature in Eq.\eqref{eq1} for the set of sample coherent states. We indeed saw that the value obtained by the nonlinear measurement matches the theoretical predictions, is unbiased both in the mean and the variance, and that the added noise is determined by the nature of the ancillary quantum state \cite{Supplement}. In addition, for comprehensive characterization of a quantum measurement, we consider a the Schr\"odinger picture.

The ideal measurement of nonlinear quadrature in Eq.\eqref{eq1} with zero excess noise projects the measured field into the nonphysical displaced cubic phase state in Eq.\eqref{eq2}.
The practical realization of Fig. \ref{fig2}B projects the field into unnormalized quantum states, which we call detector states in this paper \cite{POVM}.
This projection determines both the probability of obtaining the particular measurement result $m$ and the quantum state prepared in the case when the measurement was applied to one part of a maximally entangled state. 
The detector states can be reconstructed by detector tomography from conditional probabilities $ \mathrm{prob}(m | \alpha)$, using the set of coherent states $\{\ket{\alpha}\}$ forming an overcomplete basis for Hilbert space of the system, and the iterative maximum likelihood analysis\cite{PhysRevA.64.024102, Supplement}. 
This is also advantageous experimentally as the coherent states are resilient to losses and have been already employed for tomography of homodyne and photon number resolving detectors \cite{Lundeen2009, Grandi_2017}.

\begin{figure}
    \centering
    \includegraphics[width=.95\textwidth]{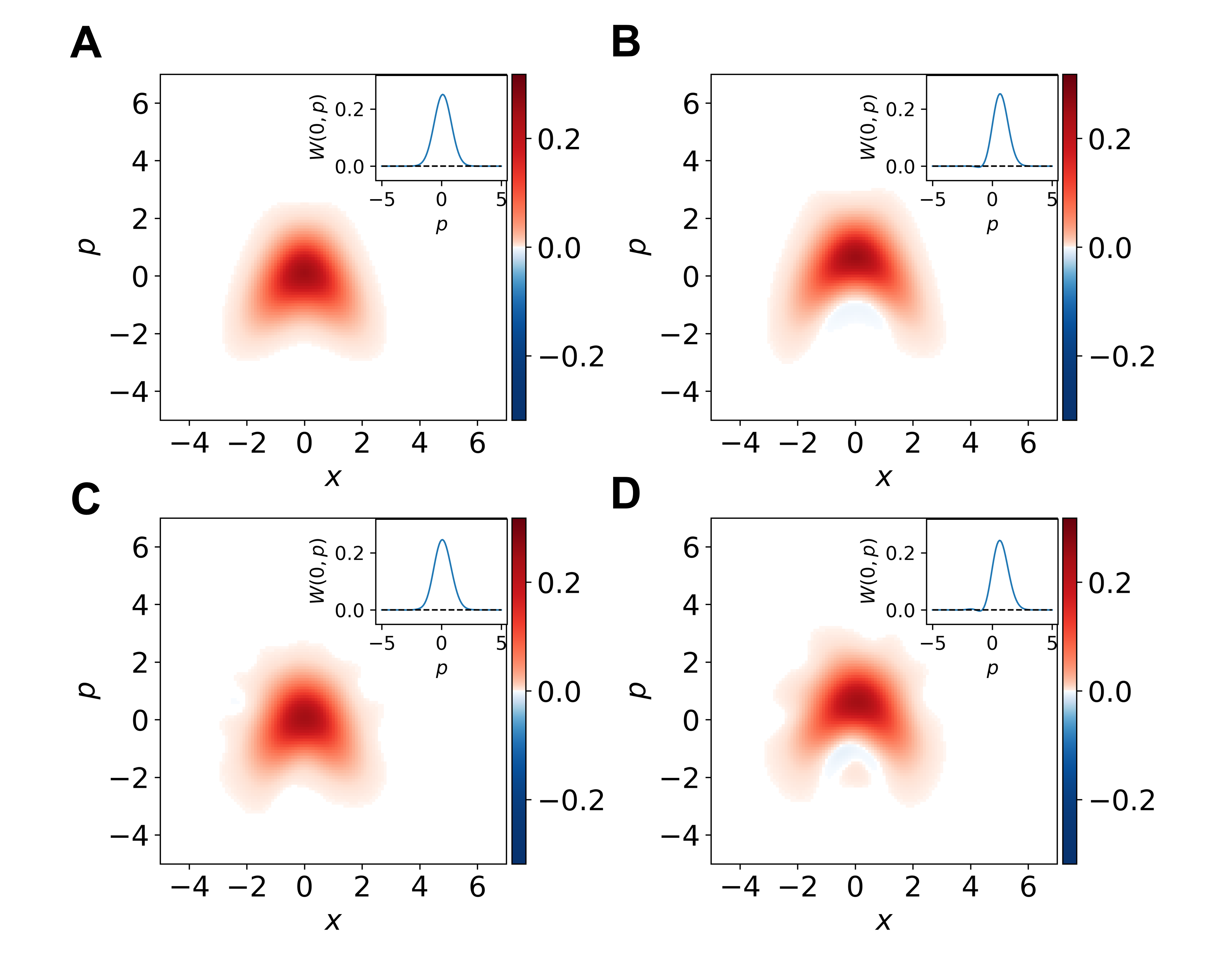}
    \caption{\textbf{Averaged detector states of the nonlinear quadrature measurement.}
    (A) Theoretical detector state for vacuum ancilla with $\mathrm{var}(\hat{p}+\gamma\hat{x}^2) = 0.64 $.
    (B) Theoretical detector state for non-Gaussian ancilla with $\mathrm{var}(\hat{p}+\gamma\hat{x}^2) = 0.56$.
    (C) Experimental detector state for vacuum ancilla with $\mathrm{var}(\hat{p}+\gamma\hat{x}^2) = 0.74 \pm 0.01$.
    (D) Experimental detector state for non-Gaussian ancilla with $\mathrm{var}(\hat{p}+\gamma\hat{x}^2) = 0.67 \pm 0.01$.
    The detector states are averaged with $p$-displacement by the measurement outcomes $m$ and renormalized.
    Note that there are ripples in (C) and (D) that are not present in (A) and (B), which are artifacts of the reconstruction method \cite{Supplement}. The insets show the cross section of Wigner functions along $x=0$.
    % However, these artifacts are insignificant in comparison to the significant negativity of (B) and (D).
    }
    \label{fig4}
\end{figure}

Figure \ref{fig4} shows the normalized Wigner functions of the detector states. Fig. \ref{fig4}A and Fig. \ref{fig4}B show the ideal detector states for the vacuum and the non-Gaussian ancilla with $\mathrm{var}(\hat{p}+\gamma\hat{x}^2) = 0.64$ and $\mathrm{var}(\hat{p}+\gamma\hat{x}^2) = 0.56$, respectively. Fig.~\ref{fig4}C and Fig.~\ref{fig4}D then show detector states for vacuum and actual non-Gaussian ancilla reconstructed from experimental data.
% \color{red}
Detector states for measured values $m$ differ simply by displacement in $p$, which is consistent with the analysis of the first moments. Although reconstructed detector states are noisy because of limited number of data points \cite{Supplement}, the averaged detector states show the significant properties of the nonlinear quadrature measurement in Fig.~\ref{fig4}. 
Note that the averaging and renormalizing are done after $p$-displacement by measured results $m$ to cancel the instrinsic displacement of the detector states.
% \color{black}
The parabolic shape is a qualitative evidence of the nonlinear quadrature measurement. It is determined by the nonlinear feedforward and therefore present for both kinds of ancillary states. The main difference is that the non-Gaussian ancilla leads to detector states with variance of the nonlinear quadrature operator $\hat{p}+\gamma\hat{x}^2$ equal to $0.67 \pm 0.01$ in Fig.\ref{fig4}D. In contrast, the vacuum state produces variance equal to $ 0.74 \pm 0.01$ (Fig.\ref{fig4}C). This means the excess noise of the measurement with non-Gaussian ancilla, constructed as superposition of the vacuum and single photon states, is already suppressed by 10$\%$ relative to the vacuum level, which is consistent with the measured nonlinear squeezing of the ancillary state. Thus we can see that even small nonlinear squeezing of the ancilla can already provide an observable effect. 
% The detector states for the non-Gaussian ancilla are also demonstrably highly non-classical \cite{Supplement}.

In conclusion, we have implemented a nonlinear quadrature measurement of $\hat{p}+\gamma\hat{x}^2$ using the nonlinear electro-optical feedforward and non-Gaussian ancillary states.
The nonlinear feedforward makes the tailored measurement classically nonlinear, while the ancillary state pushes the measurement into highly non-classical regime and determines the excess noise of the measurement. By using a non-Gaussian ancilla we have observed 10$\%$ reduction of the added noise relative to the use of vacuum ancillary state, which is consistent with the amount of nonlinear squeezing in the ancilla.
Higher reduction of the noise can be realized in the near future by a better approximation of the CPS using a superposition of higher photon number states\cite{PhysRevA.93.022301, takase2022}.
We can now simultaneously create broadband squeezed state of light beyond 1THz \cite{kashiwazaki2020,kashiwazaki2021}
and can make a broadband amplitude measurement on it with 5G
technology beyond 40GHz \cite{https://doi.org/10.48550/arxiv.2205.14061}, as well as a broadband photon-number measurement beyond 10GHz \cite{endo2021}.
Furthermore, the nonlinear feedforward presented here can be compatible with these technologies if an application specific integrated circuit (ASIC) is developed based on the FPGA board presented here.
By using such technologies we can efficiently create non-Gaussian
ancillary states with large nonlinear squeezing by heralding schemes \cite{PhysRevA.88.053816,fukui2022} even when the success rate is very low. It is because we can repeat heralding beyond 10GHz and can compensate for the very low success rate.

When supplied with such high-quality ancillary state, this nonlinear measurement can be directly used in the implementation of the deterministic non-Gaussian operations required in the universal quantum computation.
Our experiment is a key milestone for this development as it versatilely encompasses all the necessary elements for universal manipulation of the cluster states.
Furthermore, this method is extendable to multiple ancillary states case in implementation of the higher-order quantum non-Gaussianity\cite{PhysRevA.97.022329}.

Our experiment demonstrates an active, flexible, and fast nonlinear feedforward technique applicable to traveling quantum states localized in time. If the nonlinear feedforward system is combined with the cluster states \cite{doi:10.1126/science.aay2645, doi:10.1126/science.aay4354} and GKP states \cite{PhysRevA.64.012310}, all operations required for large-scale fault-tolerant universal quantum computation can be implemented in the same manner.
As such, we have demonstrated a key technology needed for optical quantum computing, bringing it closer to reality.

% Your references go at the end of the main text, and before the
% figures.  For this document we've used BibTeX, the .bib file
% scibib.bib, and the .bst file Science.bst.  The package scicite.sty
% was included to format the reference numbers according to *Science*
% style.

%BibTeX users: After compilation, comment out the following two lines and paste in
% the generated .bbl file.

% \textbf{Note: Add the following bibitem [POVM] These unnormalized detector states are also known as positive operator valued measure (POVM) elements. If possible, it could be also added a footnote.}

\bibliography{scibib}

\begin{thebibliography}{10}
\bibitem{POVM}
 These unnormalized detector states are also known as positive operator valued
  measure (POVM) elements.
\end{thebibliography}

\bibliographystyle{Science}

\section*{Acknowledgments}
% Include acknowledgments of funding, any patents pending, where raw data for the paper are deposited, etc.
This work was partly supported by Japan Science and Technology Agency (Moonshot R\&D) Grant No. JPMJMS2064, Japan Society for the Promotion of Science KAKENHI Grant No. 18H05207 and No. 21J11615, the UTokyo Foundation, and donations from Nichia Corporation.
F.H. acknowledges supports from the Forefront Physics and Mathematics Program
to Drive Transformation (FoPM)
W.A. acknowledges supports from the Research Foundation for Opto-Science and Technology.
H.Y. acknowledges the Australian Research Council Centre of Excellence for Quantum Computation and Communication Technology (Project No. CE170100012).
P.M. acknowledges Grant No. 22-08772S of the Czech Science Foundation (GACR). 
R.F. acknowledges the project 21-13265X of Czech Science Foundation and EU H2020-WIDESPREAD-2020-5 project NONGAUSS (951737) under the CSA - Coordination and Support Action.

%Here you should list the contents of your Supplementary Materials -- below is an example.
%You should include a list of Supplementary figures, Tables, and any references that appear only in the SM.
%Note that the reference numbering continues from the main text to the SM.
% In the example below, Refs. 4-10 were cited only in the SM.
% \section*{Supplementary materials}
% Materials and Methods\\
% Supplementary Text\\
% \subsection{Experimental Setups}
% \subsection{circuit design of FPGA board}
% \subsection{Flexibility of FPGA board}
% \subsection{Theory of nonlinear measurement (POVM picture)}
% \subsection{Calibration of input states}
% \subsection{Correction of residual coherent states outside target wave-packet}
% \subsection{Discussion of imperfection in experimet (loss, splitting ratio, timing jitter, phase locking-point}
% \subsection{Nonlinear squeezing evaluation of POVM elements}
% Figs. S1 to S3\\
% Tables S1 to S4\\
% References \textit{(4-10)}
\renewcommand{\thefigure}{S\arabic{figure}}
\section*{Supplementary materials}
\subsection*{Materials and Methods}
\subsubsection{Experimental Setups}
% \textcolor{red}{This section is not completed yet...}
% Figure \ref{} shows the details of experimental setup.
The light source of this experiment is a continuous-wave Ti:sapphire laser with a wavelength of \SI{860}{\nano\metre}.
The light is divided into three parts.
First, one of the beams is used to pump a second harmonic generator (SHG), which is a bow-tie shaped cavity with \SI{500}{\milli\metre} roundtrip with a periodically-poled litium niobate (PPLN) crystal inside as a nonlinear medium for SHG.
% Its conversion efficiency is about 50\% when \SI{800}{\milli\watt} carrier beam is injected to the SHG.

Second part is used for local oscillator (LO) beams of homodyne detectors, passing through two acousto-optic modulators (AOMs) and a mode cleaning cavity (MCC).
% AOMs are used only for alignment by shifting the frequency of local oscillators to check the visibility of signal mode to the local oscillators of homodyne detectors.
The output beam from the MCC is distributed to two homodyne detectors. One of the local oscillator beam is coupled to a waveguide electro-optic modulator (EOM) for phase rotation by the feedforward operation.
Displacement beam for the idler mode is also picked from this beam.

The last part is used for controlling the optical path, i.e., for cavity locking (locking beams) and for phase locking (probe beams). Frequencies and amplitudes of the control beams are controlled by AOMs.
The frequencies of each beam are differently shifted for phase locking, where we actively feedback and control the phases of light to synchronize beat notes of interference to reference signals. The frequency shifts are \SI{120}{\kilo\hertz} for the probe of idler mode, \SI{5.5}{\mega\hertz} for the probe of input beam. Locking beam of the asymmetric OPO is also detuned by \SI{1}{\mega\hertz}. %to separate the beat note in the error signals.
The modulation signals and reference signals are generated by synchronized direct-digital synthesizers.
The control beams are switched on and off periodically, which is called as sample and hold technique. In the sample phase, the control beams are turned on and we activate the feedback controls of phase and cavity locking. In the hold phase, the control beams are turned off and we deactivate the feedback, keeping the condition of the optical system. This technique is used to avoid the control beams to disturb photon detection as large fake counts and to saturate the homodyne detectors.

The setup for generating the ancillary state is the same as \cite{PhysRevApplied.15.024024}.
The OPO used in this experiment is a triangle cavity with \SI{108}{\milli\metre} round trip formed by two spherical mirrors and one plate polarizing beamsplitter (PBS). Inside the cavity, a type-II periodically-poled potassium titanyl phosphate (PPKTP) crystal with \SI{20}{\milli\metre} long is put between two spherical mirrors. One of the spherical mirrors is an output coupler with a transmittance of 14\,\%.
This OPO is called an asymmetric OPO because the cavity is single resonant in the polarization to make the wave packet shape of the signal state into an exponentially rising shape for real-time quadrature measurement\cite{PhysRevLett.116.233602}.
The pump beam is enhanced by a buildup cavity around the OPO.
The asymmetric OPO is pumped and generates a two-mode squeezed state in two orthogonal polarization. The idler mode is $s$-polarized and resonant to the OPO while signal mode is $p$-polarized and does not resonant.
The idler mode is displaced via interference of a coherent state at a beam splitter of 99\,\% reflectivity.
Frequency filters put on the idler path are designed as Fabry-Perot cavities with linewidths of \SI{140.1}{\mega\hertz} and \SI{90.9}{\mega\hertz}, respectively.
The filtered idler beam is sent to an avalanche photodiode (APD; Excelitas technologies, SPCM-AQRH-16), and the click heralds the generation of non-Gaussian ancillary state.
% in a wave packet localized in time domain and is compatible with time-domain multiplexing technique.

% The signal mode is sent to the nonlinear feedforward setup.
% Herriott cell (timing control)
 The non-Gaussian ancillary state is generated in a wave packet localized in time domain and is compatible with time-domain multiplexing technique.
Thus, we have to synchronize the feedforward system with the arrival of the wave packet accompanied by a heralding signal.
To compensate the delay of the electrical trigger of the heralding signal compared to the arrival timing of optical wave packet (which is occurred by asymmetric optical path lengths, latency of avalanche photo-diodes, and a latency of cables to from APD to the FPGA board), 8.4 meters optical delay line is put on the signal path of the ancillary state.
The delay line includes a Herriott cell, whose two spherical mirrors (whose curvature radius is R=\SI{1000}{\milli\metre}) face each other at 168.5 mm distance. One of the mirrors has a hole through it to inject and output the light.
The light is injected into the cell and go back and forth between the mirrors 16 times, and is output from the cell after about \SI{28}{\nano\second}.

The beam splitter used for interference of the input state with the ancillary state is a variable beam splitter that consists of two plate PBS and a half wave plate. The transmittance of this beam splitter is set to $50\%$ during the experiment of nonlinear measurement, while it is set to about $100\%$ during the characterization of the ancillary state.
Another delay line synchronizes the optical signal and electrical signal for nonlinear feedforward operation. This delay line is implemented by a Herriott cell with two spherical mirrors (R=\SI{1000}{\milli\metre}) at the distance of \SI{645}{\milli\metre}.
The length of optical path of the delay line is about \SI{17.3}{\metre} corresponding to \SI{64}{\nano\second}.

\subsubsection{Design of feedforward circuits}
The feedforward circuits play two roles.
One is to extract quadrature information from mesurement outcomes of a homodyne detector, and the other is to calculate a nonlinear function of the quadrature.

To extract the quadrature of a real-valued temporal mode from a homodyne measurement with a continuous-wave local oscillator, we have to integrate the measurement outcomes weighted with the mode function.
This calculation can be processed in real-time with passive system if the impulse response of the measurement devices is designed as time-reversal of a desired mode function.
Hence, this technique is called as real-time quadrature measurement and used in a few researches of quantum states\cite{PhysRevLett.116.233602, Asavanant:17, PhysRevApplied.15.024024}.
For that purpose, we construct the circuits mainly with broadband and flat frequency response components, and add a low-pass filter which determines the shape of impulse response.
While the quantum state is localized in the wave packet with the bandwidth about \SI{35}{\mega\hertz}, we use homodyne detectors with about 200-MHz flat bandwidth, DC-coupled amplifiers and offset controller with about \SI{1}{\giga\hertz} bandwidth.
The low-pass filter is the same as the one used in \cite{PhysRevApplied.15.024024}, and has the frequency response corresponded to the asymmetric OPO and the filtering cavities.
% (200MHz homodyne detector, 1GHz DC-coupled amplifiers, and an offset controller, lpf)

\begin{figure}
    \centering
    \includegraphics[width=.9\textwidth]{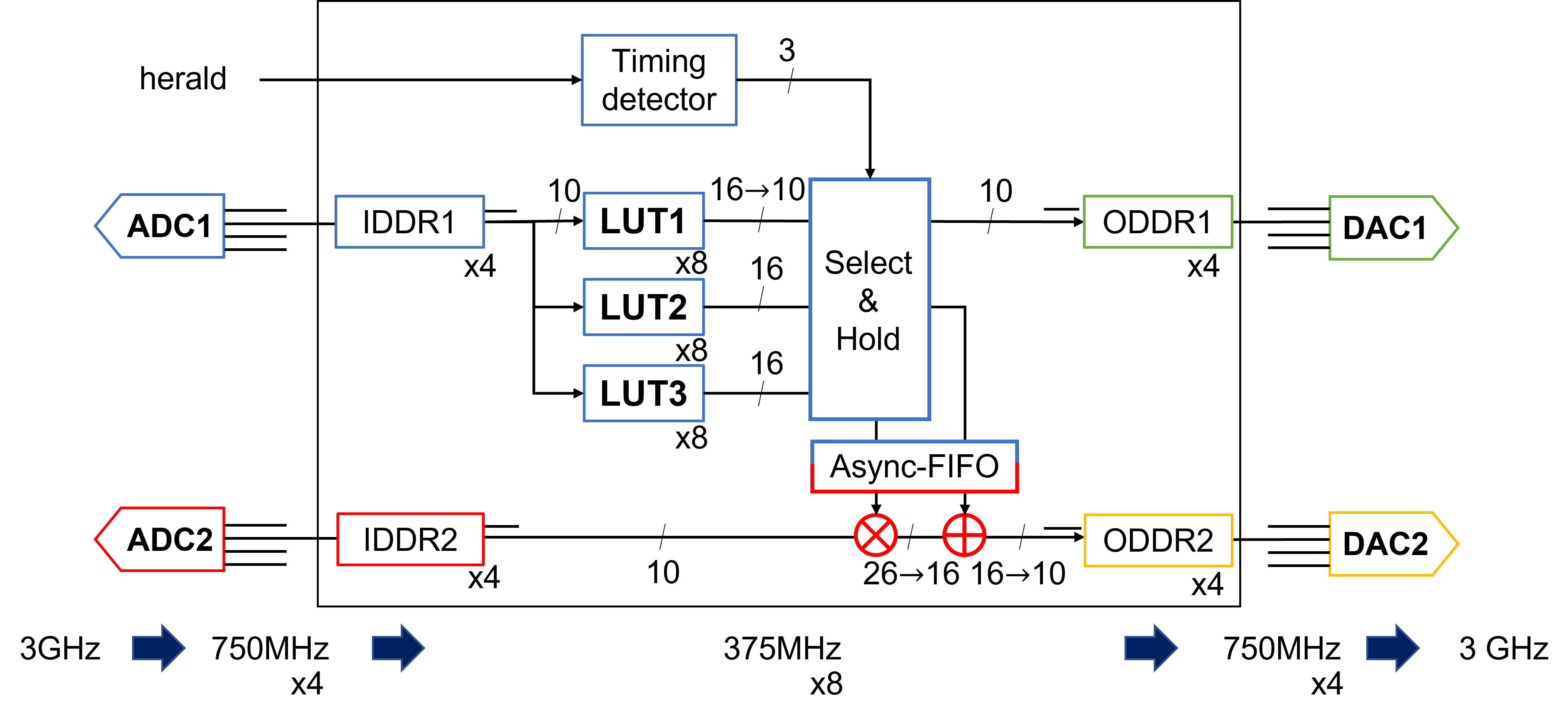}
    \caption{\textbf{Schematic diagram of the FPGA configuration.} The color of each component shows the clock domain. Numbers on the signal lines show the number of bit width. Three LUTs are implemented for future work. In this experiment, LUT1 is used for the nonlinear calculation. ADC: analog-to-digital converter, IDDR: input-double-data-rate primitive, LUT: looking-up table, ODDR: output-double-data-rate primitive, DAC: digital-to-analog converter, Async-FIFO: asynchronized fast-in-fast-out primitive.}
    \label{fig:S1}
\end{figure}
To calculate the nonlinear function of the quadrature, we employ a low-latency FPGA board (Fig.\ref{fig:S1}).
The board is equipped with two analog-to-digital converters (ADCs), two digital-to-analog converters(DACs), and a field programmable gate array (FPGA) for signal proccessing.
The ADCs (EV10AS152A, Teledyne e2v) are synchronized with a 3-GHz sampling clock. The output signal (10 bits resolution) is deserialized to 8 parallel channels inside the FPGA (Kintex-7 325T, Xilinx) with a 375-MHz processing clock via ADCs themselves and input-double-data-rate (IDDR) primitive of the FPGA.
The DACs (EV10DS130AG, Teledyne e2v) also runs at \SI{3}{\giga\hertz}, serializing the 8 parallel channels via output-DDR(ODDR) primitives of FPGA and DACs themselves. 
An essential property of the ADC and DAC is low latency, where the pipeline delay of the ADC and the DAC are 7.5 clock cycles and 4.5 clock cycles, respectively.
Analog parts of the FPGA board has about \SI{450}{\mega\hertz} $-1$dB bandwidth, which is enough broad to treat the signal from a homodyne detector.

Looking-up tables for the nonlinear calculation in the FPGA board are implemented by block random access memories (BRAMs). Pre-computed values of the nonlinear function are loaded to the BRAMs by a soft microprocessor core.
The output signal of the FPGA board is normally turned off.
A heralding signal of the ancilla preparation turns on the feedforward operation, holding the results of nonlinear calculation until the end of wave packet of the ancillary state. 
% The heralding signal is also used for the trigger of the oscilloscope for measurement.
Since the quadrature signals are deserialized to 8 parallel channels inside the FPGA, we have to take care of the timing for triggering nonlinear feedforward.
% If we just hold all of the channels, the output signal oscillates in \SI{375}{\mega\hertz}.
% If we just hold one specific channel, jitters up to \SI{2.67}{\nano\second} occur and degrade the quality of non-Gaussian ancillary state.
Because of the 8 parallel channels, if we employ conventional strategy, jitters up to \SI{2.67}{\nano\second} occur. In this work, however, we implement a time-to-digital converter for the trigger signal via a tapped delay lines, to cancel this jitter.

The output signal of the FPGA board is amplified to drive the EOM. The gain is tuned so that the range of the output voltage is the half-wave modulation voltage of the EOM to maximize the resolution of the phase rotation.

\subsubsection{Calibration of input states}
For calibrating the input coherent state, we utilize the experimental setup as a heterodyne measurement.
The heterodyne measurement of the input state is simply done if we observe the outcomes of two homodyne detectors at \SI{100}{\nano\second} before the arrival of non-Gaussian ancillary state.
At this timing, the ancillary state can be regarded as a vacuum state since we pump the OPO in weak pump condition, as well as the feedforward is deactivated to set the measured basis of two homodyne detectors to $\hat{x}$ and $\hat{p}$.
Since the coherent state rotates at \SI{5.5}{\mega\hertz} in the phase space, the exact input state can be estimated from the measurement results.
Note that we do not ignore the weak thermal state.
We measure the shot noise of the homodyne detectors, which is used to calibrate the electrical outcomes, blocking the input states and ancillary states.

% Thus, two homodyne detectors measure the $x$ and $p$ quadratures of input state with a vacuum noise.

As mentioned in the main text, we choose the amplitude from 0.0 to 3.5 by 27 steps, which is a range where the feedforward circuit is not saturated. For each amplitude, 80000 frames (including two measurement outcomes of homodyne detectors and a phase reference signal for the input coherent states) are recorded by an oscilloscope.
We fit the measured outcomes in each amplitudes to a complex amplitudes, $f(A, \phi_\mathrm{offset}) = A \exp \left[i \left(-\phi + \phi_\mathrm{offset}\right)\right] + \left(x_0+i p_0\right)$, where $A$ is the amplitude, $\phi$ is the phase of reference signal, $\phi_\mathrm{offset}$ is a phase offset from the reference, $x_0$ and $p_0$ are offset of $x$ and $p$ quadratures. $A$, $\phi_\mathrm{offset}$, $x_0$, and $y_0$ are the fitting parameters.

\begin{figure}
    \centering
    \includegraphics{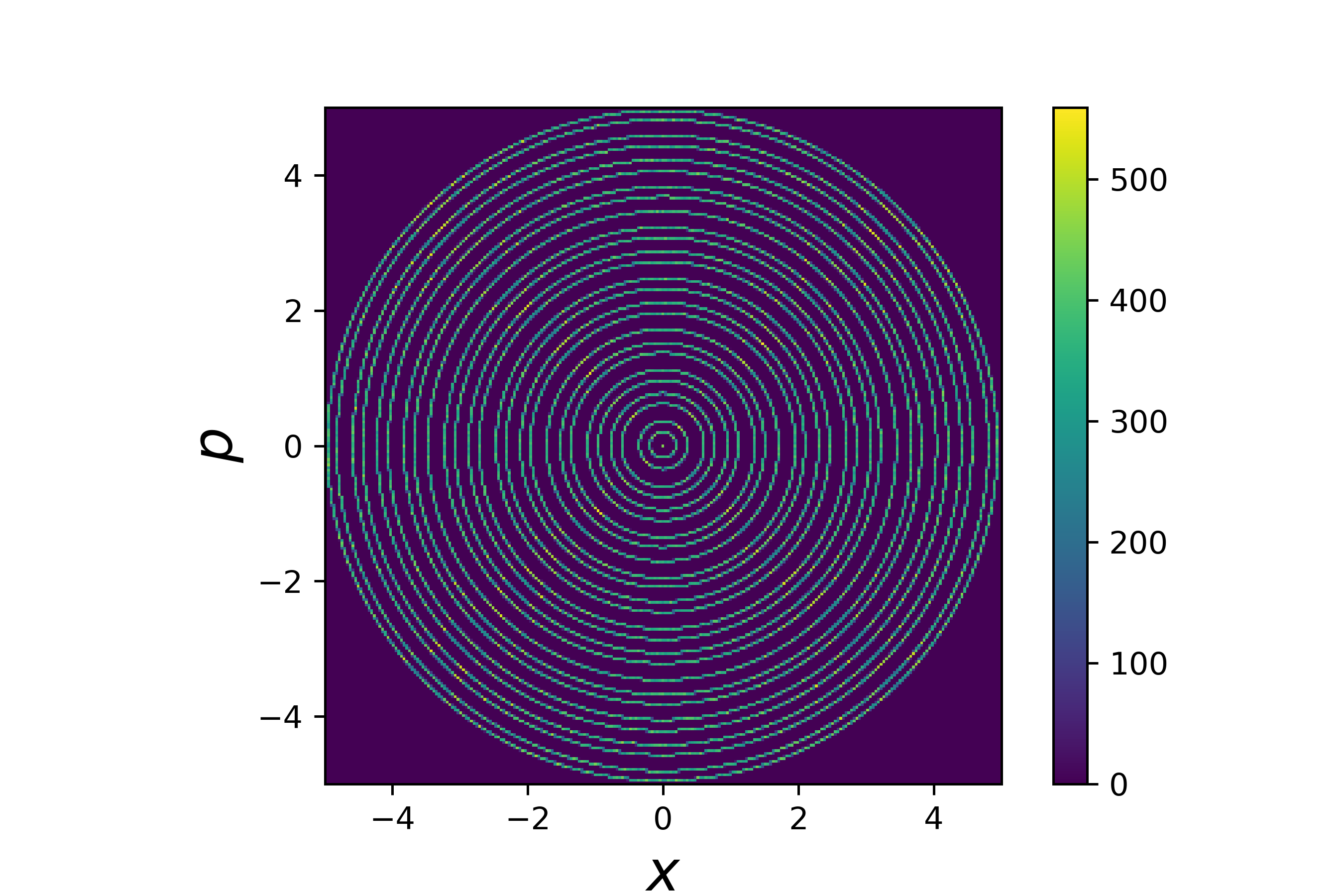}
    \caption{\textbf{Fitting of coherent states to a complex amplitudes.} The brightness of each amplitude is corrected by multiplying its amplitude, since the number of data points is inverse proportional to the amplitude and hard to see the region of large amplitude. The color bar shows corrected density of the input states and the values are not normalized.}
    \label{fig:S2}
\end{figure}
% The distribution of fitted coherent states are shown in Fig.\ref{fig:S2}. 
% In the figure, Note that the brightness of each amplitude is corrected by multiplying its amplitude, since the density of data is inverse proportional to the amplitude and hard to see the region of large amplitude.
Figure \ref{fig:S2} shows the distribution of fitted coherent states.
The fluctuation of quadrature offsets are negligibly small compared to the amplitudes of coherent states.
The amplitudes are stepped equally in enough fine resolution since the distribution of the POVM elements is derived from the ancillary states (see Eq.\eqref{eq:Pi_m}), which has no steep structures in the phase space.
The phases of the coherent states are randomized uniformly as intended.

\subsubsection{Post-processing for the measurement outcomes}
After two measurement outcomes are obtained from two homodyne detectors, we correct the effect of residual coherent state before we apply nonlinear gain $g(q)$.
This is because coherent state is injected continuously at a single frequency, while the measurement system works for a specific wave packet.
%In order to understand this correction, the timing of feedforward is important.
We perform the nonlinear feedforward to the second homodyne detector by a few nanoseconds before the arriaval of non-Gaussian ancillary state.
However, the impulse response of measurement device has a long tail in time-domain because of a high-pass filter with a cuf-off frequency of \SI{100}{\kilo\hertz} which is contained in the homodyne detectors to remove noisy low-frequency components from their electrical outcomes.
The filter has little effect when we consider the case of the non-Gaussian ancilla localized in time, because this effect is as the same as negligible loss.
When we consider the case of continuous-wave input, however, the outcomes of HD2 include information from different phase coherent states before feedforward. Thus, the effect is more significant if the amplitude of coherent state is large.

The model is explained as followed.
We define a real temporal mode function $f(t)$, which is the same mode function as the ancillary state. To consider the phase rotation by the feedforward, we consider a complex temporal mode $f(t) e^{i\theta(t)}$ where $\theta(t)$ is a rotated phase.
If we assume the nonlinear feedforward instantly rotates the phase of the measurement basis, $\theta(t)$ is a step function,
\begin{align}
    \theta(t) = 
\left\{
\begin{array}{ll}
0  & \left( t < t_\mathrm{f} \right)\\
\Theta & \left( t_\mathrm{f} \leq t \right)\\
\end{array}
\right.
\end{align}
where $t_\mathrm{f}$ is the trigger timing of the nonlinear feedforward and $\Theta$ is the rotated angle.
The contribution of coherent states in the measured value is
\begin{align}
    \int_{-\infty}^{t_\mathrm{f}} f(t) |\alpha| \sin \left(\Omega t + \phi \right) \mathrm{d}t + \int_{t_\mathrm{f}}^{\infty} f(t) |\alpha| \sin \left(\Omega t + \phi + \Theta \right) \mathrm{d}t
\label{eq:phase_rotate}
\end{align}
where $\Omega$ is angular frequency of the coherent state and $\phi$ is the phase offset of the coherent state.
Because the mode function $f(t)$ is localized around $t = 0$ even with the long tail, if the timing of nonlinear feedforward operation is sufficiently earlier than the arrival of wave packet, in other words $t_F \to -\infty$, this contribution becomes 
\begin{align}
    \int_{-\infty}^{\infty} f(t) |\alpha| \sin \left(\Omega t + \phi + \Theta \right) \mathrm{d}t
\end{align}
This is what we should measure.
Thus, the residual offset is calculated as
\begin{align}
    c(\phi, \Theta) = |\alpha| \int_{-\infty}^{t_\mathrm{f}} f(t) \left[ \sin \left(\Omega t + \phi + \Theta \right) - \sin \left(\Omega t + \phi \right)\right] \mathrm{d}t
\end{align}

We experimentally characterize the correction factor $c(\phi, \Theta)$. We input coherent states and vacuum ancillay states to the experimental setup and program the LUT to rotate the phase of the local oscillator $\Theta$ by +90 and -90 degrees.
The measured value is actually Eq.\eqref{eq:phase_rotate} but the second term can be cancelled by summing the two results with +90 and -90 degrees.
The estimated the correction factor is $c(\phi, \Theta) = 0.161 |\alpha| \left[\sin (\Omega t+ \phi +\Theta-0.812) - \sin (\Omega t+ \phi-0.812)\right]$.

Note that, to avoid this correction, preparation of coherent states in a localized wave packet is possible in principle, but it requires much longer optical delay lines to wait the preparation of input states after the heralding events because the heralding signals appear at random timings.
After the correction, the outcomes is multiplied with the nonlinear gain $g(q) = \sqrt{2} \slash \cos \theta(q) = \sqrt{1+2 \gamma^2 q^2}$ to obtain the measurement result of whole setup.

\subsubsection{First and second moments of measured nonlinear quadratures}
\begin{figure}
    \centering
    \includegraphics[width=.95\textwidth]{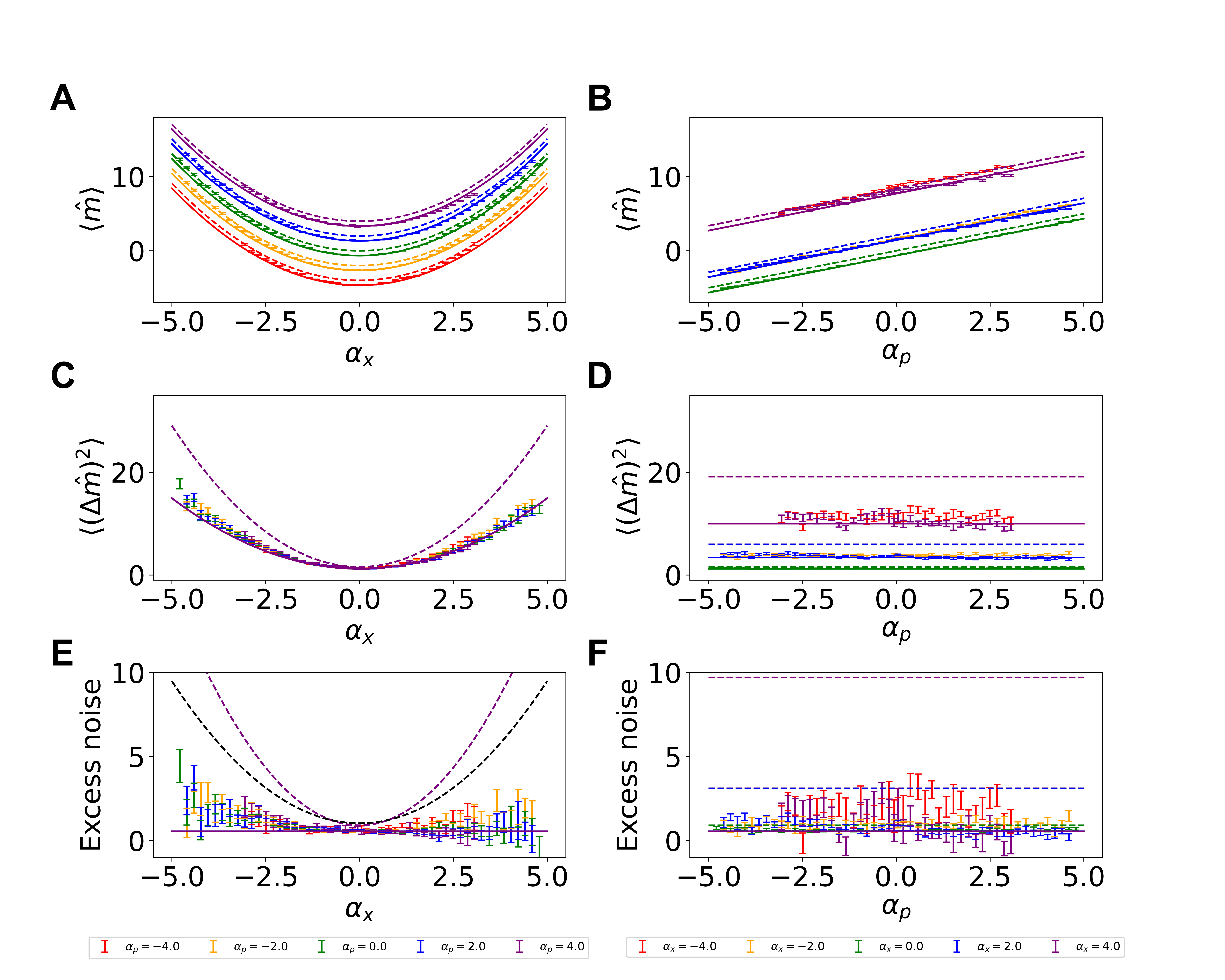}
    \caption{\textbf{Outcomes of the tailored measurement versus amplitudes of input coherent states.}
    % The amplitude of input coherent states is expressed by $\alpha=\left(\alpha_{x}+i\alpha_{p}\right) \slash \sqrt{2}$.
    % = |\alpha| e^{i\phi}$.
    (A) Mean values of the measurement outcomes $m$ as a function of $\alpha_{x}$ with fixed $\alpha_{p}$. (B) Mean values of the measurement outcomes $m$ as a function of $\alpha_{p}$ with fixed $\alpha_{x}$. (C) and (D) show the variance of the measurment outcomes $m$ as a function of $\alpha_x$ and $\alpha_p$. (E) and (F) show the excess noise of the measurement outcomes $m$ as a function of $\alpha_x$ and $\alpha_p$.
    Solid lines show the theoretical predictions calculated with the measured ancillary state.
    Dashed lines are theoretical plots without the nonlinear feedforward.
    }
    \label{fig:S3}
\end{figure}
% Fig.\ref{fig:S3} shows the input/output relation of the measurement.
% Scatter plot on $\alpha_{x}$-$m$ plane (Fig.\ref{fig:S3}A) shows the linear dependency on $\alpha_{p}$ of measurement result, while $\alpha_{p}$-$m$ plane (Fig.\ref{fig:S3}B) shows the quardratic dependency on $\alpha_{x}$, as the measured operator is $\hat{p}+\gamma\hat{x}^2$.

The nonlinear feedforward enables us to access the nonlinear quadrature of the input states, including a nonlinear quadrature term of ancillary state as followed:
\begin{align}
    \hat{m} =
\hat{p}_{\mathrm{in}} +\gamma\hat{x}^{2}_{\mathrm{in}} +
\left( \hat{p}_{\mathrm{anc}} -\gamma\hat{x}^{2}_{\mathrm{anc}} \right)
\label{eq:unbiased_meas}
\end{align}
As a simple check, we calculate first and second moments of the measurement outcomes.

Figure \ref{fig:S3} shows the statistics of the measurement outcomes $m$ as a function of the input coherent states.
We observe the measurement outcomes $m$ depend quadratically on $\alpha_x$ (Fig.\ref{fig:S3}A and \ref{fig:S3}C), and linearly on $\alpha_p$ (Fig.\ref{fig:S3}B and \ref{fig:S3}D). The experimetal mean values (Fig.\ref{fig:S3}A and \ref{fig:S3}B) also show good agreement with the theoretical predictions based on the ancillary state used in the experiment. %shown in Fig.\ref{fig2}B.
On the other hand, the experimental variances (Fig.\ref{fig:S3}C and \ref{fig:S3}D) agree with the theoretical predictions with small $\alpha_{x}$, while show a relatively large deviation when $\alpha_x$ is large.
This is because the accuracy of arctangent calculations in the nonlinear feedforward is limited for the larger $\alpha_x$ in addition to less number of data for larger $\alpha_x$ with $\alpha_p = 0$.
The deviation of the variances does not depend on $\alpha_p$ since the value is not measured by the homodyne measurement HD1 and not used in the nonlinear feedforward.

Dashed lines in Fig.\ref{fig:S3} represent theoretical lines without feedforward, highlighting the importance of the nonlinear feedforward.
Without feedforward, the system turns out to be an inadaptive linear heterodyne measurement\cite{1056132}, where $\hat{x}_{\mathrm{in}}$ and $\hat{p}_{\mathrm{in}}$ are simultaneously measured with quantum noise of the ancillary state, $\hat{x}_{\mathrm{anc}}$ and $\hat{p}_{\mathrm{anc}}$.
If the measured values of the heterodyne measurement, $q', p'$ are nonlinearly processed to calculate $\sqrt{2}p'+2\gamma q'^2$, the processed measurement result $\hat{m}_{\mathrm{het}}$ is given by
\begin{align}
\hat{m}_{\mathrm{het}}=
\hat{p}_{\mathrm{in}} +\gamma\hat{x}^{2}_{\mathrm{in}} +
\left( \hat{p}_{\mathrm{anc}} +\gamma\hat{x}^{2}_{\mathrm{anc}} \right)
-2 \hat{x}_{\mathrm{in}} \hat{x}_{\mathrm{anc}}
\label{eq3}
\end{align}
This is a kind of measurement about $\hat{p}_{\mathrm{in}}+\gamma\hat{x}^2_{\mathrm{in}}$.
In classical schemes, this setup works well since we can ignore the noise from the ancillary states.
Compared to the unbiased noise terms in Eq.\eqref{eq:unbiased_meas}, however,
the last term of Eq. \eqref{eq3} indicates that the noise term is biased by $\hat{x}_{\mathrm{in}}$ of the input, and cannot be cancelled by any ancillary states. (Eigenstates of $\hat{x}$ suppress the cross term, but $\hat{p}$ of the states completely cover the measurement results of the input states.)
Note that we use the vacuum state as the ancillary state for the dashed lines in Fig.\ref{fig:S3}.
Figure \ref{fig:S3}C and \ref{fig:S3}D show that our nonlinear feedforward eliminates this unwanted dependence of noise on $x$ quadratures of the input.
Although the variance of experimental data is not completely unbiased to the input state due to the imperfection in the experiment, the additional noise is less biased and reduced by at least 40\% from the case without feedforward.

Moreover, the advantage of our measurement over inadaptive Gaussian measurements is verified by the excess noise (Fig.\ref{fig:S3}E and Fig.\ref{fig:S3}F).
A general inadaptive Gaussian measurement can be simplified to an unbalanced heterodyne measurement without the nonlinear feedforward and it has a biased noise term when used for the measurement of $\hat{p}_{\mathrm{in}}+\gamma\hat{x}^2_{\mathrm{in}}$ by a nonlinear post-processing.
This noise term can be minimized with respect to a known set of input coherent states but it can never be completely removed.
Black dashed line shows the case of optimal Gaussian measurement minimizing the average excess noise for coherent states of the same distribution of the experimental input states. %coherent states uniformly spread on a line on $|\alpha_x| < 5 $ and $\alpha_p = 0$.
The excess noise of our nonlinear quadrature measurement is smaller than the bound of nonlinearly processed Gaussian measurements without nonlinear feedforward in all input states.
Therefore, our nonlinear quadrature measurement overcomes general Gaussian measurements via a non-Gaussianity induced by the nonlinear feedforward.

% In general, a Gaussian measurement detects the pair of operators
% \begin{align}
%     \hat{x}_\mathrm{G} = \hat{x} + \hat{x}_\mathrm{g} ,\,\,
%     \hat{p}_\mathrm{G} = \hat{p} + \hat{p}_\mathrm{g}
% \end{align}
% where $\hat{x}$ and $\hat{p}$ with $\left[\hat{x}, \hat{p} \right]=i\hbar$ are quadrature operators of measured quantum state, and $\hat{x}_\mathrm{g}$ and $\hat{p}_\mathrm{g}$ with $\left[\hat{x}_\mathrm{g}, \hat{p}_\mathrm{g} \right]=i\hbar$ represent the minimal noise that need to be added to make the measurement possible.

% This measurement is represented by POVM elements corresponding to displaced squeezed states.
% The general Gaussian measurement can be realized by an unbalanced heterodyne measurement.
% The parameters $g$ and $\theta$ can be controlled by the splitting ratio of the beamsplitter in the heterodyne measurement.
% Note that any Gaussian measurement composed of arbitrary number of Gaussian operations, Gaussian ancillary states, and homodyne detectors can be reduced to this form.
% To implement a measurement of nonlinear quadrature $\hat{p}+\gamma\hat{x}^2$, the measurement results are processed.
% \begin{align}
%     \hat{Q}_\mathrm{NL} = \hat{p}_\mathrm{G} + \gamma \hat{x}^2_\mathrm{G}
%     = \hat{p} + \gamma \hat{x}^2 + \left(\hat{x}_\mathrm{g} \hat{x} + \hat{x}^2_g \right) - \hat{p}_\mathrm{g}
% \end{align}

% Since the noise term is biased by $\hat{x}$, 

\subsubsection{Detector tomography of nonlinear quadrature measurement}
%safety range, 
For evaluation of quantum property of our measurement, we perform detector tomography of the tailored measurement and
reconstruct the detector states (POVM elements) via an iterative maximum likelihood method \cite{PhysRevA.64.024102}.
In the analysis, we limit the area of POVM elements in the phase space representation
because we can correctly reconstruct the POVM elements only within the area covered by coherent probe states.

\begin{figure}
    \centering
    \includegraphics[width=.95\textwidth]{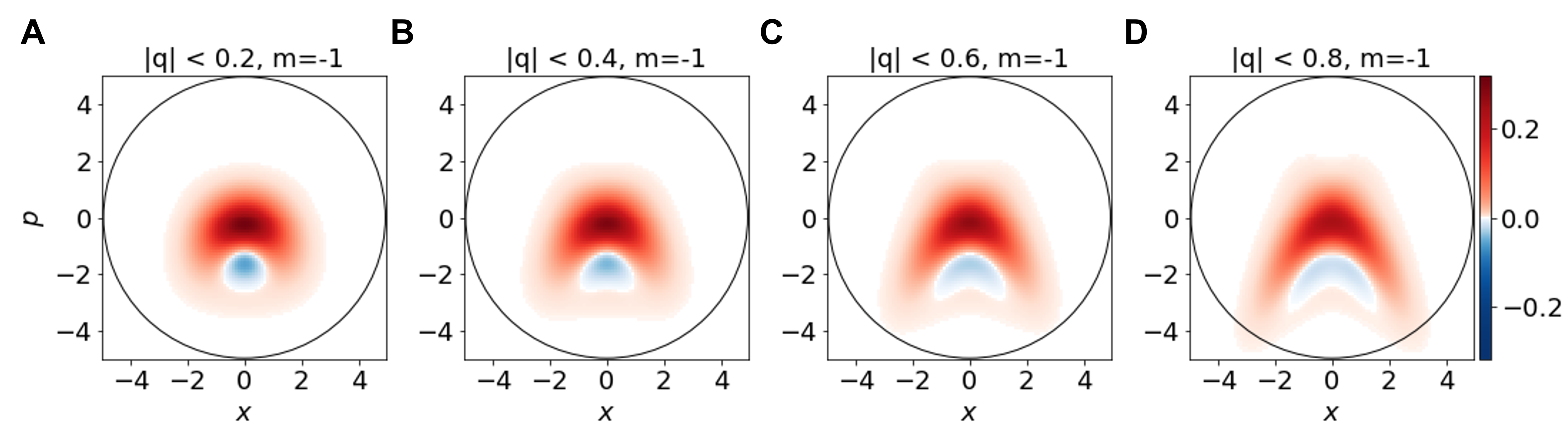}
    \caption{\textbf{Theoretical POVM elements with different integral ranges.} Wigner functions of ideal detector states associated with $m = -1$. Note that the ancillary state used for the plot is a pure state, $\ket{\psi} = 0.8\ket{0} + 0.6 i \ket{1}$. The range of $q$ is (A) $|q| < 0.2$, (B) $|q| < 0.4$, (C) $|q| < 0.6$, and (D) $|q| < 0.8$.
    Black circle shows the area covered by the input states.}
    \label{fig:S4}
\end{figure}

To confirm the area occupied by the POVM elements, we theoretically calculate the POVM elements predicted from the measured ancillary state (see Eq.\eqref{eq:Pi_m_range}).
Figure \ref{fig:S4} shows the Wigner functions of predicted POVM elements with different integral ranges of $q$ and different $m$.
Note that the each POVM element is renormalized as the trace to be 1.
The POVM elements corresponding to $-1\leq m \leq 1$ and $-0.6 \leq q \leq 0.6$ is almost inside the area scanned by the input states.

\begin{figure}
    \centering
    \includegraphics[width=.95\textwidth]{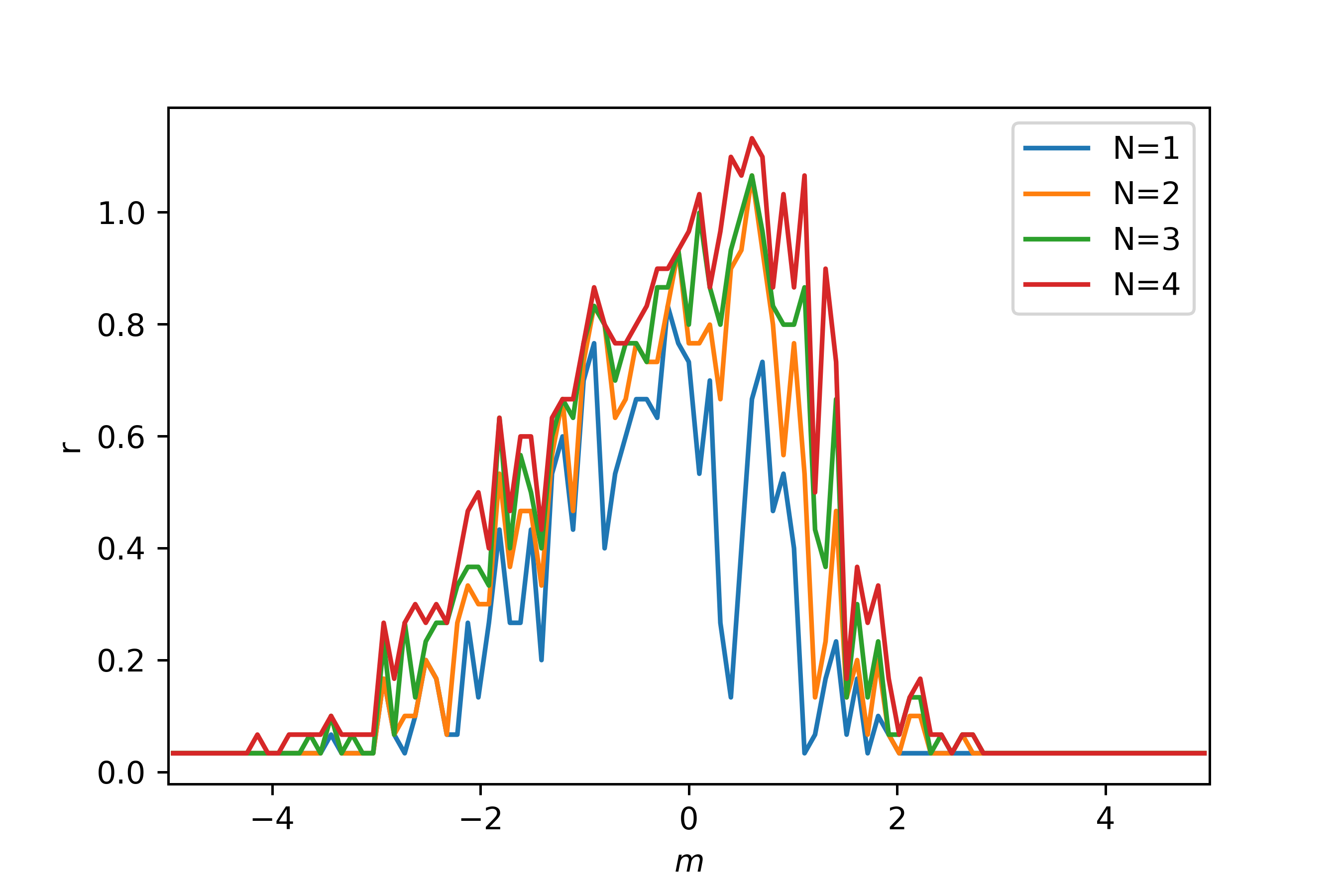}
    \caption{\textbf{Safety range of $q$ for a given $m$.} The plot shows the minimal range of $q$ ($|q| < r$) for the given $m$ including only $N$ events on the boundary coherent states.}
    \label{fig:S5}
\end{figure}

These boundaries of integral ranges are double-checked via a distribution of measurement outcomes with input coherent states on the boundary of the covered area.
The probability to obtain a certain measurement outcomes $Q, M$ is calculated as
\begin{align}
    \mathrm{prob}(Q, M | \alpha) = \mathrm{Tr} \left[ \ket{\alpha}\bra{\alpha} \hat{\Pi}_{Q,M} \right] 
    = \iint W_{\alpha} (x, p) W_{\hat{\Pi}_{Q,M}} (x, p) \mathrm{d}x \mathrm{d}p
\end{align}
Hence, if a POVM element which is inside the area covered by the coherent probe states, and if the coherent states of the maximum amplitude are injected,
the probability to obtain the measurement outcomes regarding to the POVM element will be negligibly small.
Figure \ref{fig:S5} shows the minimal range $r$ where $|q| < r$ with given $m$ include only $N$ events when $|\alpha|=3.5$.
Within $|q|<0.6$ and $-1 \leq m \leq 1$, almost no event is observed on the boundary input states.
Note that if we choose $|\alpha| < 3.5$, we have about 240,000 events within $|q|<0.6$ and $-1 \leq m \leq 1$ in total.
The measurement outcomes $m$ are distributed continuously, but for the sake of evaluation we discretize the measurement range into 20 events and reconstruct the respective detector states.

\begin{figure}
    \centering
    \includegraphics[width=.95\textwidth]{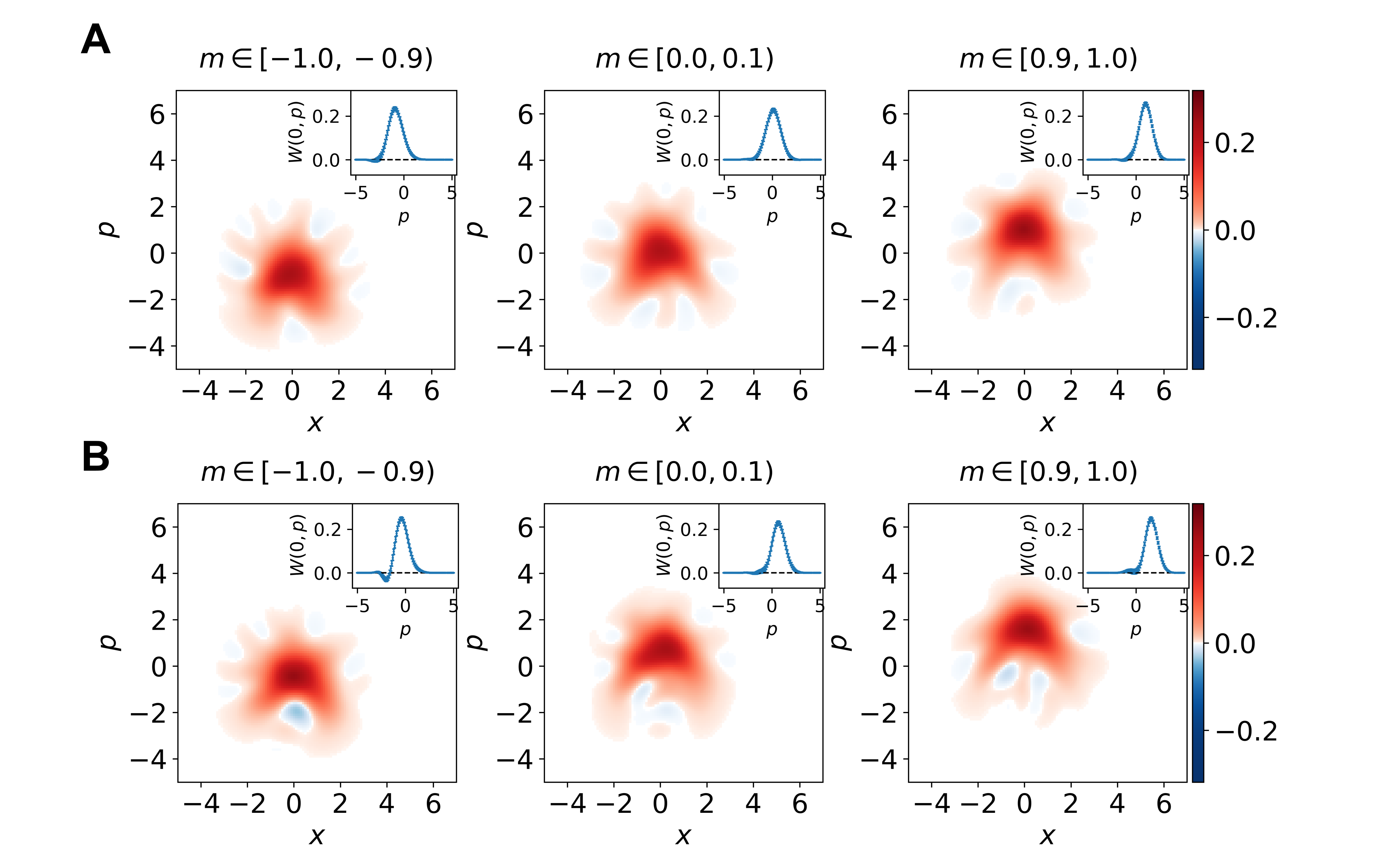}
    \caption{
    \textbf{Wigner functions of reconstructed detector states.}
    % Each POVM elements are shown in the form of conditional probabilities with the outcome in (A) $[-1.0, -0.9)$, (B) $[0.0, 0.1)$, (C) $[1.0, 1.1)$.
    % \textbf{Note for figure: move figure C to supplement, rename x axis by 'measured value'. Instead of existing A and B, plot: A) ideal POVM element for perfect ancilla, B) POVM element for vacuum, C) POVM element for NL ancilla. Pick m for which the NL ancilla has visible negativity. }
    %(A) Theoretically ideal detector state with $\mathrm{var}(\hat{p}+\gamma\hat{x}^2) = 0$.
    Wigner functions of the reconstructed detector states associated with the outcome $m$ in $[-1.0, -0.9)$, $[0.0, 0.1)$, and $[0.9, 1.0)$ injecting (A) a vacuum ancillary state, and (B) the non-Gaussian ancillary state.
    The insets show the cross section of Wigner functions along $x=0$ with error bars calculated by a bootstrapping method.
    % Each Q-function is normalized as the integration over the display range to be 1.
    % The values in the area where not covered by the input states are assumed to be zero.
    % The red dashed lines show the parabola of $p+0.52 x^2 = m+0.33$, which are consistent with the offset of the non-zero mean of $\hat{\delta}_{\mathrm{anc}}$ of the ancillary state.\textit{TO BE REPLACED}
    }
    \label{fig:S6}
\end{figure}

Figure \ref{fig:S6} shows the Wigner functions of reconstructed detector states associated with different measurement outcomes.
The detector states are displaced in $p$-direction by the measurement outcomes.
Though Wigner functions are have some ripples, the detector states have more sharp parabolic shapes with the non-Gaussian ancillary states compared to vacuum ancillary states.

\subsubsection{Artifacts in the reconstruction method}
% Negative values in the Wigner functions of quantum states are sufficient evidence of quantum non-Gaussianity of the quantum states.
% We evaluate the negativity of the Wigner functions of detector states to check the quantum non-Gaussianity of our measurement.
The area with negative values in the Wigner functions of the detector states should be induced by a negativity in the quantum non-Gaussian ancillary states, that is also induced via a two-mode entanglement and on/off detection on the idler mode.
Even in the case of vacuum ancillary states, however, there are some ripples around the Wigner functions with negative values.

\begin{figure}
    \centering
    \includegraphics[width=.95\textwidth]{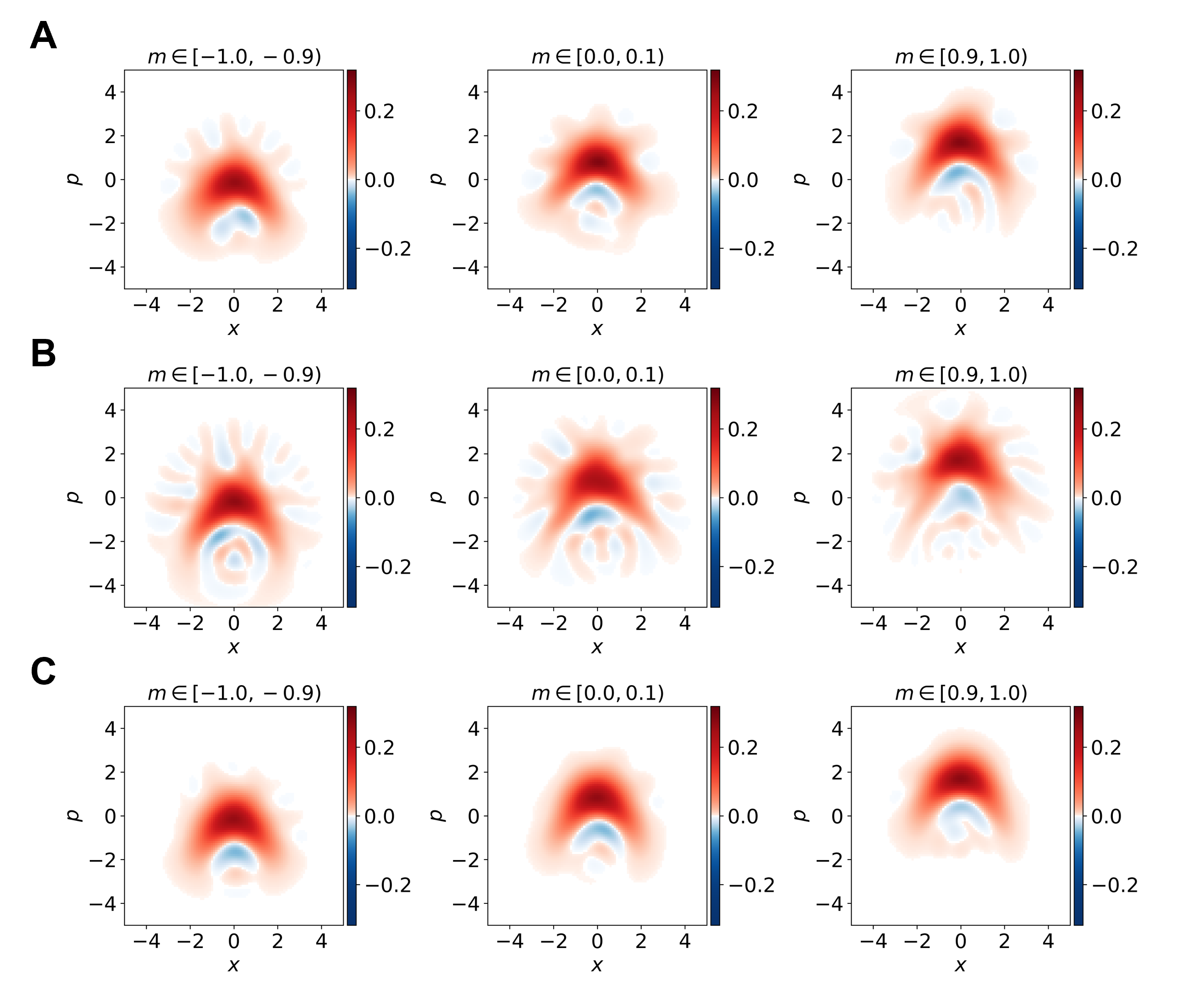}
    \caption{
    \textbf{Detector states reconstructed from Monte-Carlo simulations with different configuration.}
    Wigner functions of detector states are reconstructed with different maximum photon number $N_\mathrm{max}$ and different number of data points. 
    (A) $N_\mathrm{max}=10$ with 2.16 million points, (B) $N_\mathrm{max}=15$ with 2.16 million points, and (C) $N_\mathrm{max}=10$ with 21.6 million points.
    The data points are theoretically generated via Monte-Carlo simulation following section.\ref{sssec:simulation}, with $\eta_1 = \eta_2 = 1$, and pure non-Gaussian ancilla $\hat{\rho}_\mathrm{A} = \ket{\psi}\bra{\psi}$, $\ket{\psi} = 0.8 \ket{0} - 0.6 i \ket{1}$.
    }
    \label{fig:S7}
\end{figure}

The ripples are considered as artifacts in the reconstruction method, regarding finite photon number subspace and finite number of experimental results.
Figure \ref{fig:S7} shows the effect of these configurations in the reconstruction method.
The ripples are emphasized with more maximum photon numbers calculated in the reconstruction method.
More data points reduce the artifacts but still visible artifacts remain.
Note that more data points have less advantage in actual experiment, because much longer time will be required to obtain 10 times larger number of data and 
the entire system will become more unstable.

\subsubsection{Noise reduction by the non-Gaussian ancillary state}
\begin{figure}
    \centering
    \includegraphics[width=.95\textwidth]{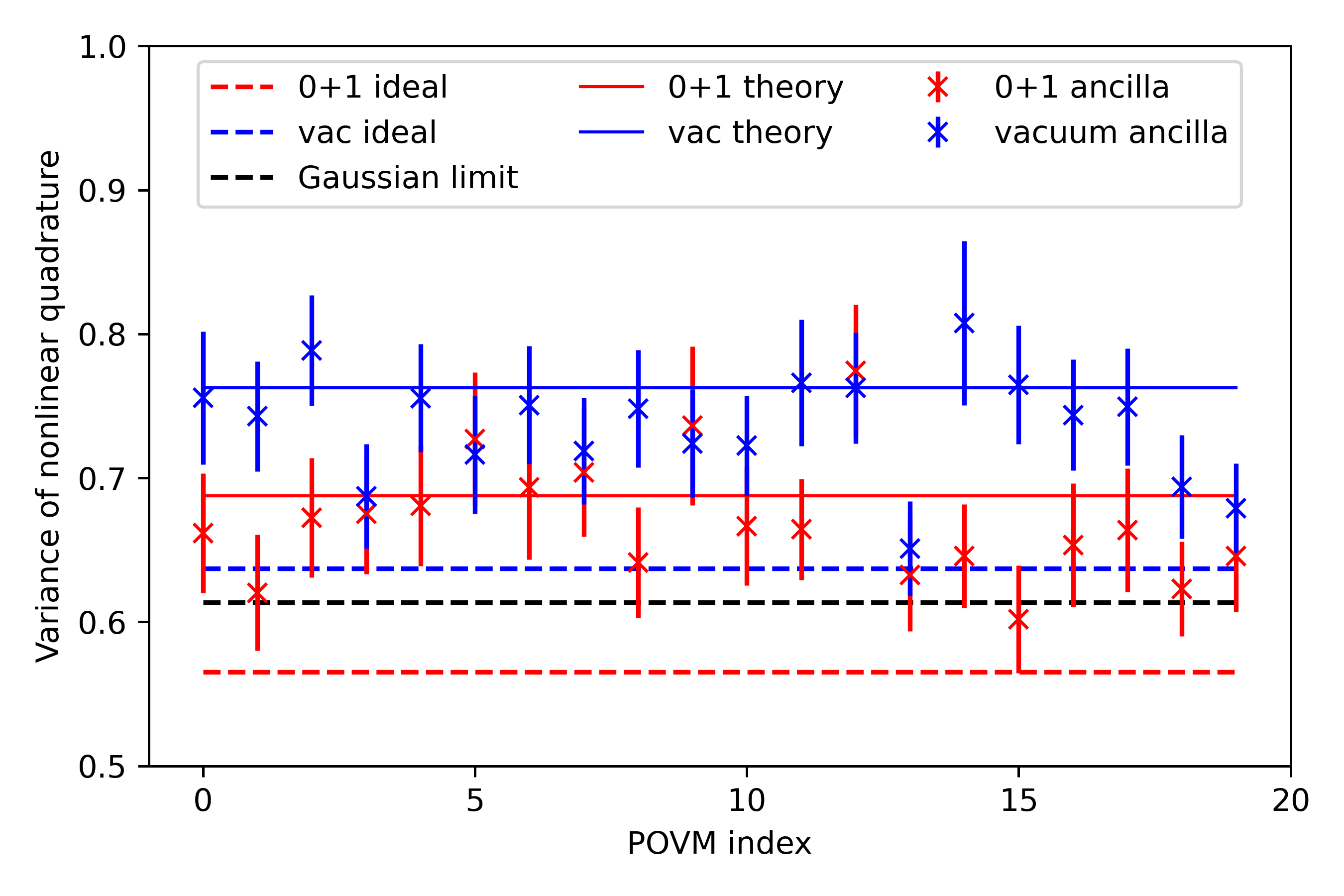}
    \caption{
    \textbf{Excess noise level of detector states.}
    $\mathrm{var}(\hat{p}+\gamma\hat{x}^2)$ of experimental detector states with different measurement results. The indexes linearly correspond the ranges of measurement outcomes $m$, where the index number 0 means the interval $[-1.0, -0.9)$, and the index number 19 means the interval $[0.9, 1.0)$. The index 20 (events collecting the result out of bounds) is omitted.
    Red and blue dashed lines show ideal values of $\mathrm{var}(\hat{p}+\gamma\hat{x}^2)$, which are equal to $\mathrm{var}(\hat{p}-\gamma\hat{x}^2)$ of the non-Gaussian ancillary states and the vacuum ancillary states.
    Black dashed line shows the lower bound of $\mathrm{var}(\hat{p}+\gamma\hat{x}^2)$ for arbitrary Gaussian states and their mixture.
    Red and blue solid lines show the expected values of $\mathrm{var}(\hat{p}+\gamma\hat{x}^2)$ with actual detection efficiencies of the two homodyne detectors, $\eta_1 = 0.97$ and $\eta_2 = 0.91$.
    }
    \label{fig:S8}
\end{figure}

The non-Gaussian ancillary state lets the detector states be nearer to that of ideal nonlinear quadrature measurement of $\hat{p}+\gamma\hat{x}^2$.
The detector states of the ideal nonlinear quadrature measurement are $p$-displaced cubic phase state (CPS), in other words, eigenstates of $\hat{p}+\gamma\hat{x}^2$.
As discussed in \cite{PhysRevApplied.15.024024}, the variance of $\hat{p}+\gamma\hat{x}^2$ is an indicator of the similarity to the displaced CPS.
Hence, the advantage of the non-Gaussian ancillary state is quantitively characterized as the reduction of the variances of $\hat{p}+\gamma\hat{x}^2$ of the normalized POVM elements, $\hat{\Pi}_m \slash \mathrm{Tr} \left[\hat{\Pi}_m\right]$.
Figure \ref{fig:S8}B shows the variances of $\hat{p}+\gamma\hat{x}^2$ of the reconstructed detector states.
The variance, $0.74 \pm 0.01$ with vacuum ancillary state in average, is decreased to $0.67 \pm 0.01$ with the non-Gaussian ancillary state, which is consistent with the expected values from the measured ancillary state, from $0.64$ to $0.56$.

The variance is larger than the ideal case, mainly due to the measurement efficiency in the experimental setups. Depending on the numerical simulation, actual experimental parameter $\eta_1 = 0.97$ and $\eta_2 = 0.91$ are consistent with the experimental results (Fig.\ref{fig:S8}B).

\subsection*{Supplementary Text}
\subsubsection{Theory of nonlinear quadrature measurement (Schr\"odinger picture)}
In this section, we derive a form of POVM elements of our nonlinear quadrature measurement and show the quantum non-Gaussianity of the POVM elements.
The theory is based on \cite{PhysRevA.93.022301}, where generalized heterodyne measurement is analyzed.

First, we define basic components.
Quadrature operators, $\hat{x}$ and $\hat{p}$, which are canonical conjugate and satisfy $\left[\hat{x}, \hat{p} \right] = i\hbar$, correspond to real part and imaginary part of complex electric field for a temporal mode.
Definition of quadrature operators has a freedom of phase $\theta$. $\hat{x}_{\theta}, \hat{p}_{\theta}$ are defined as followed,
\begin{align}
    \left(\begin{array}{c}
         \hat{x}_{\theta}  \\
         \hat{p}_{\theta}
    \end{array}\right)
    =
    \left(\begin{array}{cc}
         \cos \theta & -\sin \theta  \\
         \sin \theta & \cos \theta
    \end{array}\right)
    \left(\begin{array}{c}
         \hat{x}  \\
         \hat{p}
    \end{array}\right)
\end{align}

Eigenstates of a quadrature operator $\ket{x;\hat{x}}$, which satisfies $\hat{x}\ket{x;\hat{x}} = x \ket{x;\hat{x}}$, can be defined theoretically. 
(In general, eigenstates of an operator $\hat{A}$ belonging to an eigenvalue $a$ is described as $\ket{a;\hat{A}}$). %but not physical since infinite energy is needed for the quadrature states.
Since homodyne measurements are measurement of a quadrature in a specific phase, they project quantum states to the eigenstates of the quadrature operator. Hence, a detector state (POVM element) of the homodyne measurement associated to the outcome $q$ at a phase $\theta$ is,
\begin{align}
    \hat{\Pi}^{(\theta)}_{q} = \ket{q;\hat{x}_{\theta}}\bra{q;\hat{x}_{\theta}}
\end{align}

%BEAMSPLITTER

In heterodyne measurement, which is also called as dual homodyne detection\cite{}, the measured state is interfered with an ancillary state $\hat{\rho}_\mathrm{anc}$ at a balanced beamsplitter, then measured by two homodyne detectors whose measurement bases are set to orthogonal quadrature phases.
Probability to obtain two outcomes $q, y$ from the two homodyne detectors with the bases of $\hat{x}$ and $\hat{p}$ is expressed with a POVM element $\hat{\Pi}^\mathrm{(het)}_{q,y}$,
\begin{align}
    \mathrm{prob}\left(q,y\right) 
    &= \mathrm{Tr}_{\mathrm{in}}\left[ \hat{\rho}_{\mathrm{in}} \hat{\Pi}^{\mathrm{(het)}}_{q,y} \right] \\
    &= \mathrm{Tr} \left[ \hat{B} \left(\hat{\rho}_\mathrm{in} \otimes \hat{\rho}_\mathrm{anc}\right) \hat{B}^\dagger \ket{q;\hat{x}} \bra{q;\hat{x}} \otimes \ket{y;\hat{p}} \bra{y;\hat{p}} \right]
\end{align}
where $\hat{B}$ is an operator of the balanced beam splitter.
% where $\hat{\Pi}^{\mathrm{(het)}}_{q,y}$ is a POVM elemenent of heterodyne measurement associated to the measurement outcomes $q, y$.
Thus, when the ancillary state is a vacuum state, the POVM element is
\begin{align}
    \hat{\Pi}^{\mathrm{(het)}}_{q,y} &= \frac{1}{\pi\hbar}\ket{\psi^\mathrm{(het)}\left(q,y\right)} \bra{\psi^\mathrm{(het)} \left(q,y\right)} \\
    \ket{\psi^\mathrm{(het)}\left(q,y\right)} &= \bra{0} \hat{B}^\dagger \ket{q;\hat{x}} \ket{y;\hat{p}} = \ket{\sqrt{2}\alpha}
\end{align}
where $\alpha = \left(q+iy\right) \slash \sqrt{2\hbar}$ is a complex amplitude and $\ket{\sqrt{2} \alpha}$ is a coherent state.

In our nonlinear quadrature measurement, the heterodyne measurement is generalized by employing an adaptive nonlinear feedforward $\theta(q) = \arctan \left(\sqrt{2}\gamma q\right)$ and a non-Gaussian ancillary state.
With a pure ancillary state $\ket{\psi}_\mathrm{anc}$, the POVM element altered by the nonlinear feedforward to a basis of the second homodyne detector is,
% \color{red}
\begin{align}
    \hat{\Pi}^{\mathrm{(nqm)}}_{q,y} &= \frac{2}{|{\cos\theta(q)}|}\ket{\psi^\mathrm{(nqm)}\left(q,y\right)} \bra{\psi^\mathrm{(nqm)} \left(q,y\right)} \\
    \ket{\psi^\mathrm{(nqm)}\left(q,y\right)} 
    % &=  \bra{\psi}_\mathrm{anc} \hat{B}^\dagger \ket{q;\hat{x}} \ket{y;\hat{p}_{\theta(q)}}\\
    &= \hat{U}(q,y) \hat{T} \ket{\psi_\mathrm{anc}} \\
    \hat{U}(q, y) &= \hat{P} \left(\tan\theta(q) \right) \hat{D}\left(\sqrt{2}q, \frac{\sqrt{2}y}{\cos\theta(q)} 
    - \sqrt{2}q\tan\theta(q)\right)
\end{align}
where $\hat{P}(k) = \exp \left[i k \hat{x}^2 \slash \hbar \right]$ is a shear operation, $\hat{D}$ is a displacement operation and $\hat{T}$ is an anti-unitary operation, which transforms $\hat{x} \to \hat{x}$ and  $\hat{p} \to -\hat{p}$. The anti-unitary operator is derived from the bra of ancillary state.
The details of the calculations are described in \cite{PhysRevA.93.022301}.
Another POVM element $\hat{\Pi}^{\mathrm{(nqm)}}_{q,y'}$, which is associated with different measurement outcome $y'$, is just displaced in the $p$-direction from $\hat{\Pi}^{\mathrm{(nqm)}}_{q,y}$ because $y$ in $U(q,y)$ decides only the amount of $p$-displacement.
% \begin{align}
%     \hat{\Pi}^{\mathrm{(nqm)}}_{q,y'} = \hat{D}(0, y'- y) \, \hat{\Pi}^{\mathrm{(nqm)}}_{q,y} \, \hat{D}^\dagger(0, y'- y)
% \end{align}

When the ancillary state is an ideal cubic phase state $\Ket{\mathrm{CPS}} = \ket{0; \hat{p}-\gamma \hat{x}^2}$,
\begin{align}
    \hat{U}(q,y) \hat{T} \ket{0; \hat{p}-\gamma \hat{x}^2} 
    = \Ket{\frac{\sqrt{2}y}{\cos\theta(q)}; \hat{p} + \gamma \hat{x}^2}
\end{align}
Hence, this measurement is a projective measurement of $\hat{p} + \gamma \hat{x}^2$, with the measurement outcomes $m = \sqrt{2} y \slash \cos \theta(q)$. Intuitively, the shear operation recovers $p$-axial symmetry of a parabolic shape of the cubic phase state, which is broken by the displacement operation. 

The operations $\hat{U}(q,y) \hat{T}$ transforms the nonlinear quadrature operator to a similar nonlinear quadrature operator.
% \textcolor{red}{Is this expression correct?}
\begin{align}
    \hat{U}(q,y) \hat{T} \left(\hat{p} + \gamma \hat{x}^2\right) \hat{T}^\dagger \hat{U}^\dagger(q,y) = -\left(\hat{p} - \gamma \hat{x}^2\right) - \frac{\sqrt{2} y}{ \cos\theta(q) }
\end{align}
Note that the sign of the coefficient $\gamma$ is flipped by the anti-unitary operation. This means that the variance of nonlinear quadrature $\hat{p} - \gamma \hat{x}^2$ of the projected state is as same as that of $\hat{p} + \gamma \hat{x}^2$ of the ancillary state.
% Because of this relation, even when the ancillary state is not a cubic phase state, the variance of $\hat{p} + \gamma \hat{x}^2$ of the projected states is preserved as that of the ancillary state.
Hence, the variance of $\hat{p} + \gamma \hat{x}^2$ of a normalized POVM element $\hat{\Pi}^\mathrm{(nqm)}_{q,y} \slash \mathrm{Tr}\left[\hat{\Pi}^\mathrm{(nqm)}_{q,y}\right]$ is preserved even when the ancillary state is a mixed state.
This means that the nonlinear squeezing of the ancillary state is transferred to the POVM elements by our nonlinear feedforward.

If we focus on the measurement outcomes of our nonlinear quadrature measurement, $m = \sqrt{2}y \slash \cos \theta(q)$, the POVM element $\hat{\Pi}_{m}$ is represented by a simple integration of $\hat{\Pi}^{\mathrm{(nqm)}}_{q,y}$ with respect to $q$, because only one real $y$ always exists for given $m$ and arbitrary real $q$.
\begin{align}
    \hat{\Pi}_m = \int_{-\infty}^{\infty} \mathrm{d}q \, \hat{\Pi}^\mathrm{(nqm)}_{q, \sqrt{2}m\slash\cos\theta(q)}
\label{eq:Pi_m}
\end{align}
Since the POVM elements $\hat{\Pi}_{q,y}$ for all $q$ and $y$ preserve the nonlinear quadrature from the ancilla, even though the sign is flipped, $\hat{\Pi}_{m}$ also keeps the same nonlinear squeezing as $\hat{\Pi}_{q,y}$.

In actual experiment, we consider a finite interval of integral $[-r, r]$ since the range of the probe coherent states is limited by the experimental constraint.  
\begin{align}
    \hat{\Pi}_m^\mathrm{(exp)} = \int_{-r}^{r} \mathrm{d}q \, \hat{\Pi}^\mathrm{(nqm)}_{q, \sqrt{2}m\slash\cos\theta(q)}
\label{eq:Pi_m_range}
\end{align}
The parameter $r$ is determined by the range of the amplitude of the coherent states and the POVM elements.

\subsubsection{Nonlinear quadrature measurement with experimental imperfection}
\label{sssec:simulation}
In this section, we consider the model of our nonlinear quadrature measurement including experimental imperfection (optical losses and phase fluctuations).
If we assume the efficiency is equivalent to linear loss, a detector state of a homodyne detection with the detection efficiency $\eta$ is modeled as followed,
\begin{align}
    \hat{\Pi}_\eta (q | \theta) &= \int \ket{tq+ry;\hat{x}_\theta}\bra{tq+ry;\hat{x}_\theta} |\braket{0;\hat{n} | -rq+ty;\hat{x}_\theta}|^2 \mathrm{d}y \\
    t &= \sqrt{\eta}, r = \sqrt{1-\eta}
\end{align}
where $q$ is the measurement outcome and $\theta$ is the measurement basis.

With this imcomplete homodyne detector, the whole setup of our adaptive measurment is modeled as,
\begin{align}
    \hat{\Pi}(q, y) = \mathrm{Tr}_{\mathrm{A}}\left[ \hat{B} \left( \hat{\Pi}_{\eta_1}(q | 0) \otimes \hat{\Pi}_{\eta_2} (y | \theta(q)) \right) \hat{B}^\dagger \hat{\rho}_{\mathrm{A}}\right]
    \label{eq:povm_imperfection}
\end{align}
where $q, y$ are associated measurement outcomes, $\theta(q) = \arctan \left(\sqrt{2} \gamma q\right) + \frac{\pi}{2}$ is the rotated angle determined by the nonlinear feedforward, $\eta_1$ and $\eta_2$ are efficiencies of two homodyne detectors including propagation losses after the beam splitter, and $\hat{\rho}_\mathrm{A}$ is the ancillary state.

% For your review copy (i.e., the file you initially send in for
% evaluation), you can use the {figure} environment and the
% \includegraphics command to stream your figures into the text, placing
% all figures at the end.  For the final, revised manuscript for
% acceptance and production, however, PostScript or other graphics
% should not be streamed into your compliled file.  Instead, set
% captions as simple paragraphs (with a \noindent tag), setting them
% off from the rest of the text with a \clearpage as shown  below, and
% submit figures as separate files according to the Art Department's
% instructions.

\end{document}